\documentclass[fleqn,usenatbib]{mnras}

\usepackage{newtxtext,newtxmath}

\usepackage[T1]{fontenc}

\DeclareRobustCommand{\VAN}[3]{#2}
\let\VANthebibliography\thebibliography
\def\thebibliography{\DeclareRobustCommand{\VAN}[3]{##3}\VANthebibliography}

\usepackage{graphicx}	
\usepackage{amsmath}	
\usepackage{lscape}
\usepackage{subcaption}
\usepackage{afterpage}

\title[HI in M 106 with FAST]{Deep HI Mapping of M 106 Group with FAST}

\author[Yao Liu et al.]{
Yao Liu,$^{1,2,3,4}$\thanks{liuyao1@pku.edu.cn}
Ming Zhu, $^{1,2,3,5}$\thanks{mz@nao.cas.cn}  
Hai-Yang Yu,$^{1,2,3}$
Rui-Lei Zhou,$^{1,2,3}$
Jin-Long Xu,$^{1,2,3,5}$
Mei Ai,$^{1,2,3,5}$ 
\newauthor
Peng Jiang,$^{1,2,5}$
Li-Xia Yuan,$^{6}$
and
Hai-Yan Zhang $^{1,2}$
\\
$^{1}$National Astronomical Observatories, Chinese Academy of Sciences, 20A Datun Road, Chaoyang District, Beijing 100101, China\\
$^{2}$CAS FAST key laboratory, NAOC, Chinese Academy of Sciences, Beijing 100101, China\\
$^{3}$University of Chinese Academy of Sciences, Beijing 100049, China\\
$^{4}$Qiannan Normal University for Nationalities, Longshan Road, Tuyun 558000, China\\
$^{5}$Guizhou Radio Astronomical Observatory, Guizhou University, Guiyang 550000, China\\
$^{6}$Purple Mountain Observatory and Key Laboratory of Radio Astronomy, Chinese Academy of Sciences, 10 Yuanhua Road, Qixia District, Nanjing 210033, China\\
}

\date{Accepted 2024 October 9. Received 2024 September 30; in original form 2024 February 8}

\pubyear{2023}

\begin{document}
\label{firstpage}
\pagerange{\pageref{firstpage}--\pageref{lastpage}}
\maketitle

\begin{abstract}
We used FAST to conduct deep HI imaging of the entire M 106 group region, and have discovered a few new HI filaments and clouds. Three HI clouds/filaments are found in a region connecting DDO 120 and NGC 4288, indicating an interaction between these two galaxies. The HI features in this region suggest that DDO 120 is probably the origin of the HI stream extending from the northern end of NGC 4288 to M 106. This structure is similar to the SMC-LMC stream, but much longer, about 190 kpc. Furthermore, based on the distance measurements,
we have determined the satellite galaxy members of M 106. With an absolute magnitude cutoff of $M_{B}$ =-10, we obtained  a sample of 11 member satellite galaxies for M 106. Using the observed HI mass with FAST,
we studied the properties of satellite galaxies in M 106 and found that satellite galaxies with lower stellar masses exhibit more significant deviations from the star-forming main sequence (SFMS) in their specific star formation rates. Furthermore, the relationship between the HI mass of satellite galaxies and optical diameter generally follows the field galaxies relation. 
We discuss the possible mechanisms leading to the quenching in the M 106 group based on the new data from FAST.

\end{abstract}

\begin{keywords}
galaxies: individual: NGC 4258 – galaxies: interactions – galaxies: structure.
\end{keywords}



\section{Introduction}

M 106 (NGC 4258) is a barred spiral galaxy with absolute B-band magnitude $M_{B}=-21.20$. The galaxy has a maximum rotation velocity of 208 km s$^{-1}$ (\citealt{Erickson1999ApJ}).
Its stellar mass ($\log M_{\ast} = 10.94$ in solar units) and HI gas content ($\log M_{\text{HI}} = 9.64$ in solar units) are comparable to those of the Milky Way \citep{Karachentsev2013}.
Thus M 106 is considered to be a MW analog. The distance measured by water maser is $D=7.60\pm 0.23$ Mpc (\citealt{Humphreys_2013}), which is in reasonable agreement with the distances measured with other methods such as using Cepheid variables (e.g., \citealt{Kanbur2003A&A}; \citealt{Bono2008ApJ}), the TRGB method (e.g., \citealt{Mager2008ApJ}; \citealt{Jacobs2009AJ}) and the Tully-Fisher relation (\citealt{Tully2008ApJ}). M 106 is the brightest galaxy of the CVn II group (\citealt{Fouque1992A&AS}) and has more bright member satellites than the MW. 
It is surrounded by 19 group members whose “tidal index” $\Theta _{1} > 0$.
According to \cite{Karachentsev2014AJ}, tidal index was defined as follows:
\begin{equation}
    \Theta _{1} =\max \left [ \log_{}{\left ( L_{n}/D_{n}^{3}   \right ) }  \right ] +  C,\ \ \ \ \  n=1,2,...N.                                   \tag{1}
\end{equation}
Here $L_{n}$ is the K-band luminosity of the neighbouring n-th galaxy, $D_{n}$  is its spatial distance from the galaxy in question. The constant $C=- 10.96$ was chosen in such a way that the galaxy with $\Theta _{1} = 0$ was located at the "zero velocity sphere" relative to its main galaxy. In other words, a galaxy with $\Theta _{1} > 0$ is considered to be causally connected with its main disturber (MD) since the crossing time for this galaxy is shorter than the age of the universe $1/H_{0} $, where $H_{0}$ is the Hubble constant. Compared to the other 19 most populated suites in the Local Volume (D < 12 Mpc) listed in Table 2 in \cite{Karachentsev2014AJ}, M 106 is generally in a median-rich environment.
Meanwhile, \cite{Spencer2014ApJ} suggested that most of the probable satellites in M 106 region are blue, late-type galaxies. Besides, M 106 has more blue star-forming satellites within 240 kpc projected distance than the Local Group. As a result, satellites in M 106 probably haven’t experienced quenching, which might imply that the M 106 group is a younger system compared to the MW. Recently, deep HI imaging with the Five-hundred-meter Aperture Spherical radio Telescope (FAST) of M 106 was presented by \cite{Zhu2021ApJ}, where the discovery of a possible accretion stream toward M 106 was reported. This HI stream, with a total HI mass of $1.7\times 10^{8}M_{\odot } $ similar to that of the Magellanic stream, seems to have originated from a satellite galaxy NGC 4288. NGC 4288 is similar to the LMC in that it is a barred spiral galaxy, with an absolute B-band magnitude$M_{B} = -16.40$, a stellar mass $logM_{star}=8.61$ and a dynamic mass $logM_{26}=9.75$ in solar units (\citealt{Karachentsev2013}). Thus, this system consists of M 106, NGC 4288, and the HI steam is comparable to the MW-LMC system. 

In this work, we have mapped the entire M 106 group region with FAST to search for more diffuse gas that could be available for accretion. Our survey discovered new HI clouds, which indicate the interaction between NGC 4288 and DDO 120. This result shows that NGC 4288 and DDO 120 could be a LMC-SMC analog. 
The MW–LMC–SMC analogs are quite rare. \cite{Boylan-Kolchin2011MNRAS} concluded that there is less than a 10$\%$ chance that a MW halo with $M_{vir}=10^{12}M_{\odot }$ will host two galaxies as bright as the Magellanic Clouds while the latest estimation of MW dark matter halo mass is about $1-1.2\times 10^{12}M_{\odot }$ (\citealt{Rodriguez2022MNRAS}). Following studies searching for MW–LMC–SMC analogs in different galaxy catalogs have confirmed the results in \cite{Boylan-Kolchin2011MNRAS}. For example, \cite{Liu2011ApJ} searched for MW–LMC–SMC analogs in the Sloan Digital Sky Survey(SDSS) catalog, and they found that there is only 3.5$\%$ probability that a galaxy with MW-like luminosity will have both LMC and SMC-like satellites within 150 kpc projected distance. \cite{Robotham2012MNRAS} analyzed all Galaxy And Mass Assembly (GAMA) galaxies within a factor of 2 (±0.3 dex) of the stellar mass of the MW and found MW-type galaxies with two close companions at least as massive as the SMC are rare at the 3.4 $\%$ level (13/414, 0.01 < z < 0.055). Only two full analogs to the MW–LMC–SMC system were found in GAMA (all galaxies late-type and star-forming).
Thus it is of interest to investigate the M 106 group, as the satellite galaxies and HI contents in this system reveal a unique group environment.
We have also detected new HI filaments or High-velocity Clouds (HVCs) which have not been detected by the very deep Hydrogen Accretion in LOcal GAlaxieS Survey (HALOGAS, \citealt{Heald2011};\citealt{Kamphuis2022A&A}). 

This paper is organized as follows: section 2 presents the data collection and reduction process, including the satellite galaxy candidates selection, FAST HI observations, and data processing. Section 3 shows the results from FAST with emphasis on the new features found by FAST. In the same section, we also study the HI gas properties for satellite galaxies in the M 106 group. Section 4 provides a discussion of the origin of the HI steam based on new features found and then we proposed a possible evolution scenario for the NGC 4288-DDO 120 system. We also discussed the results of HI gas properties for M 106.
Section 5 summarizes the main results of this study.

\section{DATA COLLECTION AND DATA REDUCTIONS}

\subsection{Determining satellites for M 106}

We produce a list of potential satellites for M 106 by combining the satellite galaxy candidates catalog in \cite{Spencer2014ApJ} and \cite{Carlsten2020ApJ}, with five additional candidates drawn from \cite{Karachentsev2007AstL, Karachentsev2015AstBu, Karachentsev2020Ap} and \cite{Cohen2018ApJ}. \cite{Spencer2014ApJ} presented a catalog of 47 dwarf galaxies surrounding M 106 based on the spectroscopic observations from the Apache Point Observatory 3.5m telescope and SDSS spectra. They categorized their sample into four classes based on distance measurements and velocities: "Probable Satellites", "Possible Satellites", "Unknown if Satellites", and "Not Satellites". Meanwhile, \cite{Carlsten2020ApJ} presented a sample of 155 dwarf satellite candidates around 10 primary host galaxies in the LV based on CFHT/MegaCam imaging data. Further in \cite{Carlsten2021ApJ}, they estimated the distances of these galaxies with the surface brightness fluctuation (SBF) measurements to confirm their status as satellites and classified the galaxies into three classes: "confirmed", "possible" (unconstrained) and "background". 
The total number of potential satellites we collected for M 106 is 64. In the following section, we will discuss in detail how we derived the primary satellite catalog for M 106 from this list.

To determine the primary satellite galaxies for M 106, we first removed satellite candidates which are rejected either by \cite{Spencer2014ApJ} or \cite{Carlsten2021ApJ}. These removed galaxies are confirmed to be background/foreground galaxies according to their measured distances. The remaining subsample contains 27 candidates. Detailed notes on individual galaxies are presented in Appendix A.

Secondly, we applied a minimum luminosity cut to this 27 subsample for two reasons. One reason is that the brightness of the satellite candidates in M 106 varies greatly. For example, the apparent r-band magnitude is 12.14 for NGC 4144, while it’s 23.38 for S 14. Fainter satellites have lower galaxy mass and are supposed to have less influence on the whole system compared to the brightest satellites. The other reason is that the purpose of our work is to investigate the neutral hydrogen gas component within the M 106 galaxy group. Galaxies fainter than S 14 ($M_{B} > -10.08$) have no HI detection with FAST.
Thus we only consider galaxies brighter than $M_{B}=-10.0$ in our subsequent analysis. The apparent and absolute B-band magnitudes (corrected for galactic extinction) for each galaxy in our M 106 satellites catalog are drawn from the Hyper-Leda database (\citealt{Paturel2003A&A}), which is available on the website http://leda.univ-lyon1.fr. This selection criterion produces a sample of 11 satellite candidates for M 106. 

Thirdly, We examined all existing distance data in the literature and online database for our 11 M 106 subsample, including the LEDA and the NASA/IPAC Extragalactic Database (NED). Most satellite candidates for M 106 in our sample have duplicate distance measurements including SBF, TF, and TRGB. For galaxies located at 7 Mpc, the SBF and TF methods have the least distance accuracy, which usually have an error of roughly 1-2 Mpc, while the TRGB method generally has an error of about 0.1-0.5 Mpc. Thus we favor the more recent distance measurements by the TRGB method when there are duplicated distance values for our galaxy sample. The distances, methods of distance measurement, and references of the 11 potential satellites for M 106 are listed in Table~\ref{tableA2}.
We then used distance data for these satellite candidates combined with their projected distance to the host galaxy M 106 to categorize them into different classes with a code from 1-4 (listed in Table~\ref{tableA2}) which indicates the possibility of being a true satellite. Detailed discussions on the classification are presented in Appendix B.

Note that 16 satellite candidates for M 106 were excluded from our final satellite sample due to their significantly low luminosity.
In fact, among these satellite candidates, three galaxies (S 06, S 11, and S 16) have SBF distances similar to the distance to M 106(\citealt{Carlsten2021ApJ}), confirming them to be real satellites. And S 06 is also confirmed by \cite{Cohen2018ApJ} based on TRGB and SBF distances. The galaxy
dw1210+4727 is considered a companion of M 106 by \cite{Kamphuis2022A&A}.
Four galaxies (dw1218+4623, dw1220+4748, dw1223+4848, and dw1220+4922) have SBF distances with large errors and are classified as "possible" satellites. The SBF distance to LVJ1215+4732 is 11.7 Mpc (\citealt{Cohen2018ApJ}), much more distant than M 106. Thus we suggest it is a background galaxy. We do not conclude the membership of the remaining 7 galaxies since they have no distance or velocity information. These 16 satellite candidates for M 106 are 
good candidates but most of them need further work to confirm. The properties for them are presented in Table~\ref{tableA1}.

\subsection{HI Observation and Data Reduction for M 106}

The HI observational data for M 106 is derived from the FAST All Sky HI Survey (FASHI), which covers the northern hemisphere ($60^{\circ} > Dec > -10^{\circ} $) in a velocity range of -2000-20000 km s$^{-1}$ (\citealt{Zhang2024}). This survey employs a 19-beam receiving system (\citealt{Jiang2019, jiang2020fundamental}), structured in a hexagonal array with dual-polarization capabilities, spanning the frequency range of 1050 MHz to 1450 MHz. For the backend, the Spec(W) spectrometer was chosen, featuring 65536 channels, with each polarization and beam covering a bandwidth of 500 MHz. The velocity interval is 1.67 km s$^{-1}$, yielding a Hanning-smoothed spectral resolution of 4.8 km s$^{-1}$. The FAST HI survey is conducted through a drift scan mode, with a rotation of 23.4 deg for the 19-beam receiver, ensuring uniform beam spacing in declination at an interval of 1.14 arcmins. The FASHI will ultimately cover the entire northern FAST observation sky area at least twice to obtain complete sensitivity. A total of 12 hours of drift scan observations were conducted between September 25, 2020, and July 2021, revealing the presence of a long HI accretion stream toward M 106 (\citealt{Zhu2021ApJ}). Subsequently,  deep scan observations of the M 106 filament regions were carried out using the Multibeam On-The-Fly (OTF) mode from August to October 2021, totaling 9 hours, to confirm the weak HI sources. The OTF observations share a nearly identical configuration with the drift scan observations, with the exception that we employed a scanning separation of 10.3 arcmins and a scanning speed of 15 arcsecs per second. The system temperature for all beam/polarization channels ranges from 18 K to 22 K. At 1.4 GHz, the half-power beam width (HPBW) for each beam is approximately 2.9'. The pointing accuracy of the telescope is around 10'' (rms).

The flux calibration is performed by injecting a continuous 10K calibration signal (CAL) for 1 second every 32 seconds to calibrate the antenna temperature. The data were reduced using the HI Pipeline reduction software developed by \cite{Jing2024arX}, which includes tasks such as flux calibration, bandpass, baseline correction, radio frequency interference (RFI) detection and subtraction, spectral smoothing, and velocity reference frame conversion. Baseline correction is performed using the asymmetrically reweighted penalized least squares algorithm (arPLS, \citealt{Baek2015}).
Once the spectra are adequately calibrated, they are applied to a grid with 1' spacing in the image plane, and data cubes are created in the standard FITS format. A more detailed description of the process can be found in \cite{Xu2021ApJ}. The rms brightness temperature sensitivity is approximately 0.4 mJy/beam or 6.4 mK per channel, equivalent to a column density sensitivity of $1\sigma=5.6\times 10^{16}$ atoms cm$^{-2}$ per channel. 

\subsection{Physical parameters of satellite candidates}

We use the satellite galaxies’ velocity data and morphological data from the LEDA. They gave the Hubble type using the RC3 morphological classification(\citealt{deVaucouleurs1991}).
The stellar mass and indicative (dynamic) mass of satellite galaxies are extracted from Table 1 in \cite{Karachentsev2014AJ}
, which used Ks-band luminosity to derive the galaxy's stellar mass according to an empirical formula. The dynamic mass is derived from the formula:
\begin{equation}
  log(\frac{M_{26}}{M_{\odot}})=2log{V_{m}} +logA_{26} +logD+4.52   \tag{2}
\end{equation}
where $V_{m}$ is the rotation velocity expressed in  km s$^{-1}$, $A_{26}$ is the linear diameter of the galaxy expressed in angular minutes, determined at Holmberg’s isophote (26.5 mag arcsec$^{-2}$), and D is the distance in Mpc.
We obtain the satellite galaxies’ star formation rates (SFR) from \cite{KAR13SFR}. Such SFR values are based on the $H\alpha$ or FUV flux measurements separately. 
The physical parameters of 11 satellites for M 106 is presented in Table~\ref{tableA2}. Given that the error in distance measurements typically exceeds the virial radius, we apply a uniform distance of 7.60 Mpc for all group members when calculating distance-related parameters. Additionally, corrections have been made to the stellar mass, SFRs, and dynamical mass to account for this distance.


\section{Results}
\begin{figure*}
   \centering
  \includegraphics[width=1\textwidth] {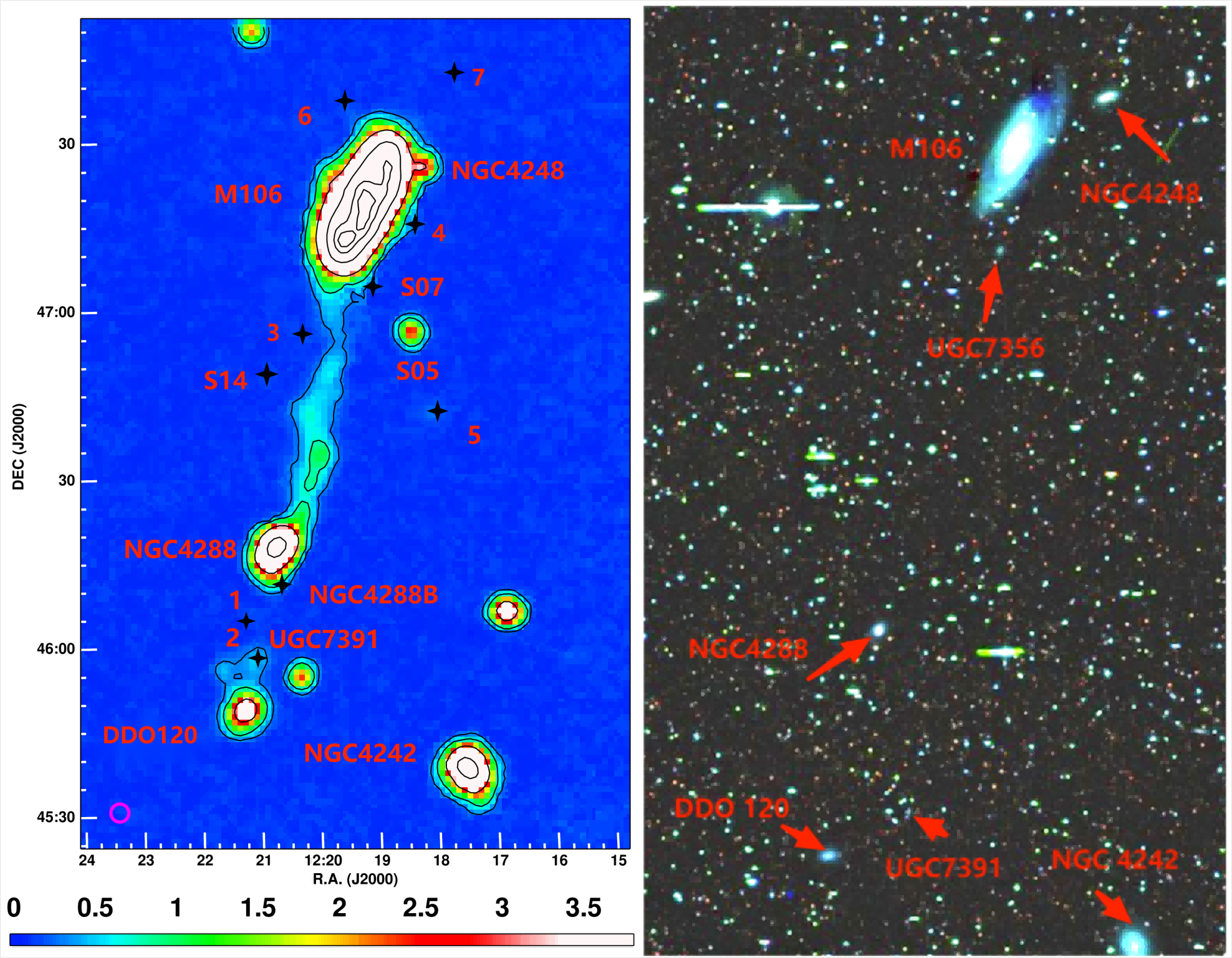}
   \caption{Left: The H I integrated intensity map of M 106 region. The velocity is in
the range of 200-805 km s$^{-1}$. The contour levels are 00.25, 0.7, 3, 15, 35, 46 Jy/beam km s$^{-1}$. The unit on the color map is Jy/beam km s$^{-1}$. The pink circle indicates the FAST beam size of 2.9'. The black asterisks indicate galaxies and HI filaments/clouds detected by FAST but not shown in this image, where the HI clouds or filaments are represented by numbers. Right: The DECals optical image for M 106 from the www.legacysurvey.org archive. 
   }
   \label{Fig1}
   \end{figure*}

\begin{figure*}
   \centering
   \includegraphics[width=0.88\textwidth] {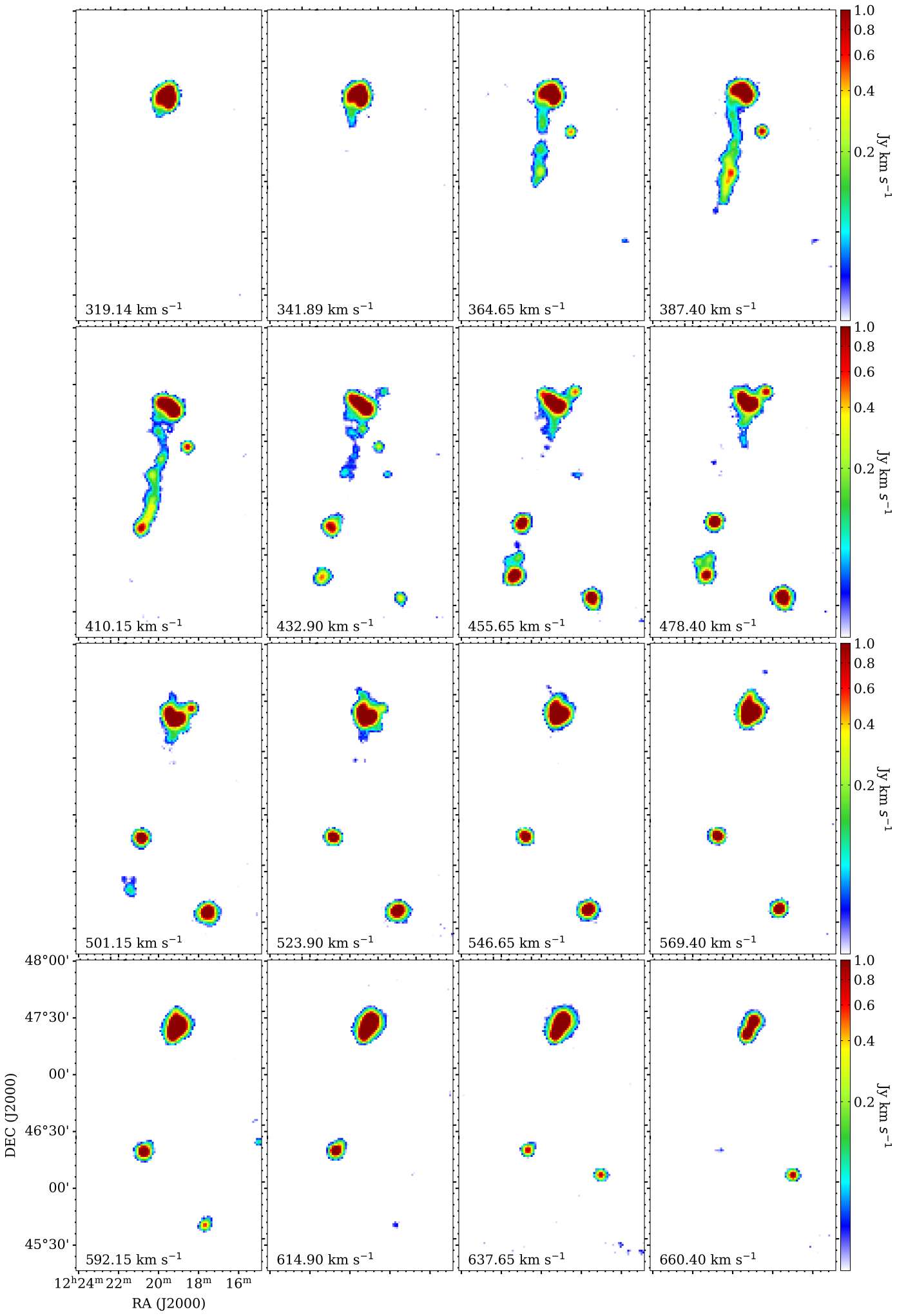}
   \caption{The channel map of the FAST H I data cube for M 106. Each channel is integrated over a velocity range of 22.75 km s$^{-1}$.
   }
   \label{Fig2}
\end{figure*}

\begin{figure}
   \centering
   \includegraphics[width=0.98\columnwidth] {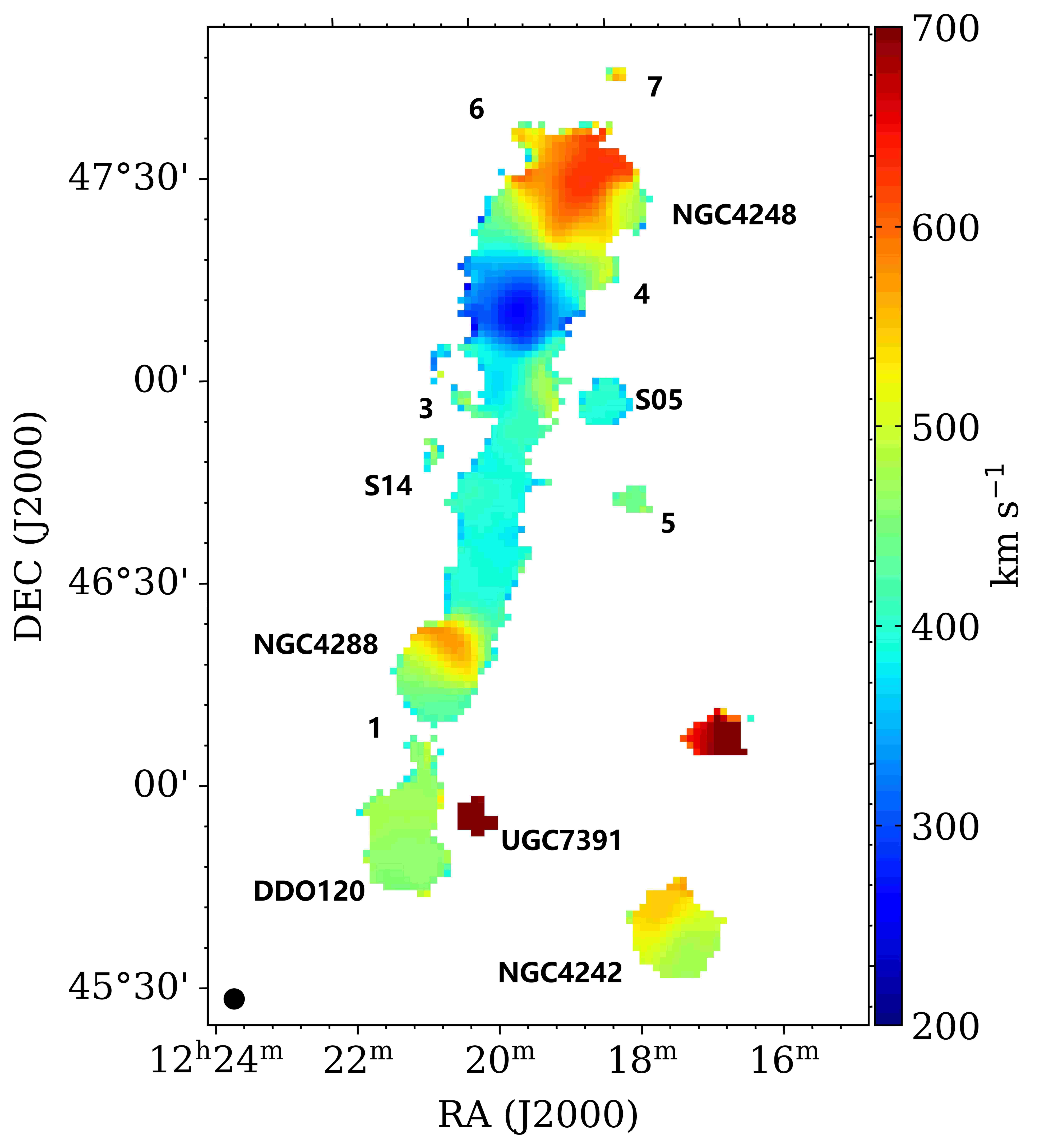}
   \caption{The moment-1 map of the FAST H I data cube for M 106.
The black filled circle indicates a beam size of 2.9 arcmin. Some HI features have been labelled.
   }
   \label{Fig3}
\end{figure}

The high sensitivity of FAST enables us to observe the HI gas structures in the M 106 region more comprehensively and achieve a more thorough measurement of the HI mass of satellite galaxies.
Figure~\ref{Fig1} shows the FAST H I integrated intensity map of the M 106 galaxy group (left panel) along with the DECals optical image (right panel). It depicts the HI emissions from M 106, NGC 4248, S 05, UGG 7356, NGC 4288, NGC 4242, DDO 120, and UGC 7391. The integrated velocity range for HI flux is 200-805 km s$^{-1}$. J1215+4730, J1211+4739, and NGC 4144 lie on the western side of M 106, with their right ascension coordinates being less than 184 degrees. Due to the coordinate range, these three galaxies are not shown in Figure~\ref{Fig1}, but are displayed separately. In Figure~\ref{FigB2}, we show the HI integrated contours overlaid on a DESI-LS optical image for J1215+4730 and J1211+4739, together with their spectral profiles taken at the peak column density. The detailed discussions of these satellite galaxies are presented in section 3.4. Furthermore, in Figure~\ref{Fig1}, we have marked all FAST-observed HI sources with black asterisks and presented their relevant parameters in Table~\ref{tableA2} and Table~\ref{tab1}. The channel map in Figure~\ref{Fig2} shows more details of the HI clouds and filamentary structures. Figure~\ref{Fig3} is the moment-1 map from FAST, which shows the velocity structure of the whole system. The observed HI sources are also shown in this figure.

\subsection{Small filaments near the M 106 disk}

The total HI mass of M 106 observed by FAST is $6.6 \times 10^{9}M_{\odot }$, which is larger than that from HALOGAS measurement( $5.7 \times 10^{9}M_{\odot }$, \citealt{Kamphuis2022A&A}).
The paper of \cite{Kamphuis2022A&A} listed some HI clouds around the M 106 disk detected by HALOGAS. With the high sensitivity, FAST also detected more HI filaments/clouds associated with the M 106 disk. Table~\ref{tab1} presents the HI clouds detected by FAST within the M 106 galaxy group. In Figure~\ref{Fig1}, we have marked their positions with black asterisks, with numbers corresponding to their identifiers. To the north of NGC 4248, there exists a cloud resembling an HVC (Cloud 7, RA=12h17m48s, Dec=$47^\circ44'04"$), with a central velocity of 575 km s$^{-1}$. FAST measurements indicate a flux of 0.065 Jy km s$^{-1}$ for this cloud, equivalent to an HI mass of $8.8\times 10^{5}M_{\odot }$.
There are also some extended emission outside the west end of the disk, below NGC4248 (Cloud 4). The total flux of this HI emission measured by FAST is 0.11 Jy km s$^{-1}$. HALOGAS also detected this HI cloud, with an HI flux of 0.106 Jy km s$^{-1}$ and a central velocity of 514 km s$^{-1}$, although no optical counterpart has been identified.

\begin{figure}
   \centering
   \includegraphics[width=0.98\columnwidth] {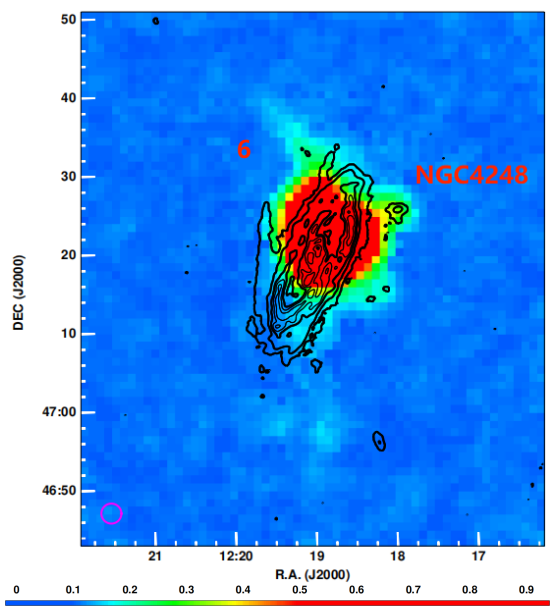}
   \caption{The HI integrated intensity map of M 106 observed by FAST with the HALOGAS contours overlaid on it. The integrated velocity range is from 516 to 550 km s$^{-1}$. The unit on the color map is Jy/beam km s$^{-1}$. Black contours are HALOGAS images (integrated over 200-750 km s$^{-1}$). The contour levels start from $1\times 10^{19}$ atoms/cm$^2$ and increase by $5.41\times 10^{20}$ atoms/cm$^2$ for each subsequent contour. Cloud 6 in the northern region of NGC 4258/M 106 is distinctly visible.
   }
   \label{Fig4}
\end{figure}

Beside the southern HI stream, there are two gas accretion streams connecting to the southeast and northeast end of M 106 disk respectively. 
These two streams are not resolved by FAST, but they can be seen clearly in the HALOGAS map (\citealt{Kamphuis2022A&A}).In Figure~\ref{Fig4} we overlaid the contours of the HALOGAS image to show the position of these two streams.
We found a new filament or HVC cloud (Cloud 6) near RA=12h19m11.7s, Dec=$47^\circ35'44.6"$, which is more extended compared to the features seen by HALOGAS (see Figure~\ref{Fig4}). It starts to show up in the channel map from 501.15 km s$^{-1}$ and ends at 546.65 km s$^{-1}$ with a central velocity of approximately 522 km s$^{-1}$. FAST measurements indicate an HI flux of 0.15 Jy km s$^{-1}$.
Figure~\ref{Fig4} shows the HI intensity map integrated over 516- 550 km s$^{-1}$. As shown in this figure, Cloud 6 has gas distribution extending in a northeasterly direction, which is perpendicular to the extending direction of the HI steam located at the northeast end of the disk.
These two tail-like structures nearly 
emerge from the same position 
on the disk. 
 They could be a result of tidal interactions between the disk of M 106 and neighboring galaxies.

\begin{table}
\centering
\caption{HI clouds observed by FAST in the M 106 group}
\label{tab1}
\resizebox{\columnwidth}{!}{%
\begin{tabular}{ccccccc}
\hline
\hline
(1)       & (2)      & (3)      & (4)         & (5)            & (6)          \\
Name      & RA       & Dec      & V           & $f_{HI}$       & $logM_{HI}$  \\
          & (J2000)  & (J2000)  & km s$^{-1}$ & Jy km s$^{-1}$ & $M_{\odot }$ \\ \hline
HI stream &          &          &             & 12.49          & 8.23         \\
Cloud1    & 12:21:15 & 46:05:00 & 464         & 0.11           & 6.18         \\
Cloud2    & 12:21:02 & 45:58:01 & 468         & 1.4            & 7.28         \\
Cloud3    & 12:20:19 & 46:56:03 & 471         & 0.052          & 5.83         \\
Cloud4    & 12:18:13 & 47:13:22 & 514         & 0.11           & 6.18         \\
Cloud5    & 12:17:52 & 46:40:43 & 446         & 0.17           & 6.36         \\
Cloud6    & 12:19:12 & 47:35:45 & 522         & 0.15           & 6.31         \\
Cloud7    & 12:17:48 & 47:44:04 & 575         & 0.065          & 5.95         \\
N4288B    & 12:20:43 & 46:12:30 & 425         & 2.00           & 7.43         \\ \hline
\end{tabular}%
}
\vspace*{3ex}
   \begin{minipage}{\columnwidth}
NOTE—Column (1): Name of the HI clouds. Column (2): R.A.(J2000). Column (3): Dec.(J2000). Column (4): Central velocity of the HI spectral line. Column (5): HI flux in Jy km s$^{-1}$ observed by FAST. Column (6): The logarithm of the HI mass in solar units derived from HI flux, assuming that HI clouds are at the same distances as M 106.
    \end{minipage}
\end{table}

\subsection{HI clouds between NGC 4288 and DDO 120}

In comparison to \cite{Zhu2021ApJ}, 
Figure~\ref{Fig1} 
covers a larger area of the M 106 galaxy group, extending both to the north and south. Additionally, it reaches deeper detection limits in certain regions. In the central area, the FAST telescope achieves a single-beam velocity resolution of 5 km s$^{-1}$ with a rms noise of 0.5 mJy/beam. Among the candidate satellite galaxies, NGC 4288 is the brightest member ($M_{B}=-18.90$), positioned at a projected distance of 139 kpc to the south of M 106. The total integrated flux measured by FAST is 47.6 Jy km s$^{-1}$, corresponding to an HI mass of $6.48\times 10^{8}M_{\odot }$. This galaxy has also been detected in the Westerbork Synthesis Radio Telescope (WSRT) CVn Survey (\# 62, \citealt{KovaC2009}). Compared to the WSRT observation (34.27 Jy km s$^{-1}$), FAST detected 38.9 $\%$ more flux.

\begin{figure*}
    \centering
    \begin{subfigure}[b]{0.95\textwidth}
      \includegraphics[width=\textwidth]{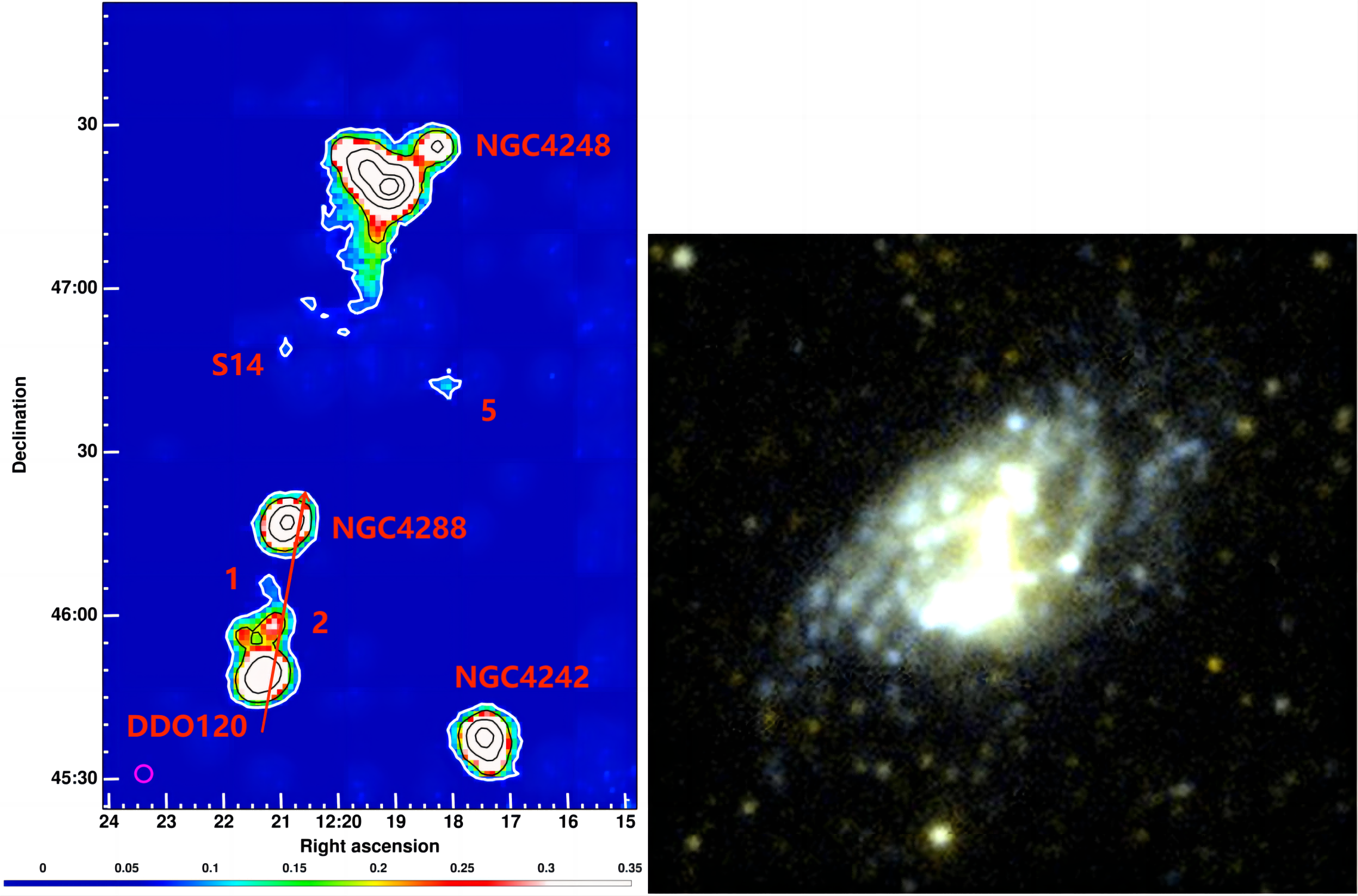}
      \caption{}
    \end{subfigure}%
    \\
    \begin{subfigure}[b]{0.55\textwidth}
      \includegraphics[width=\textwidth]{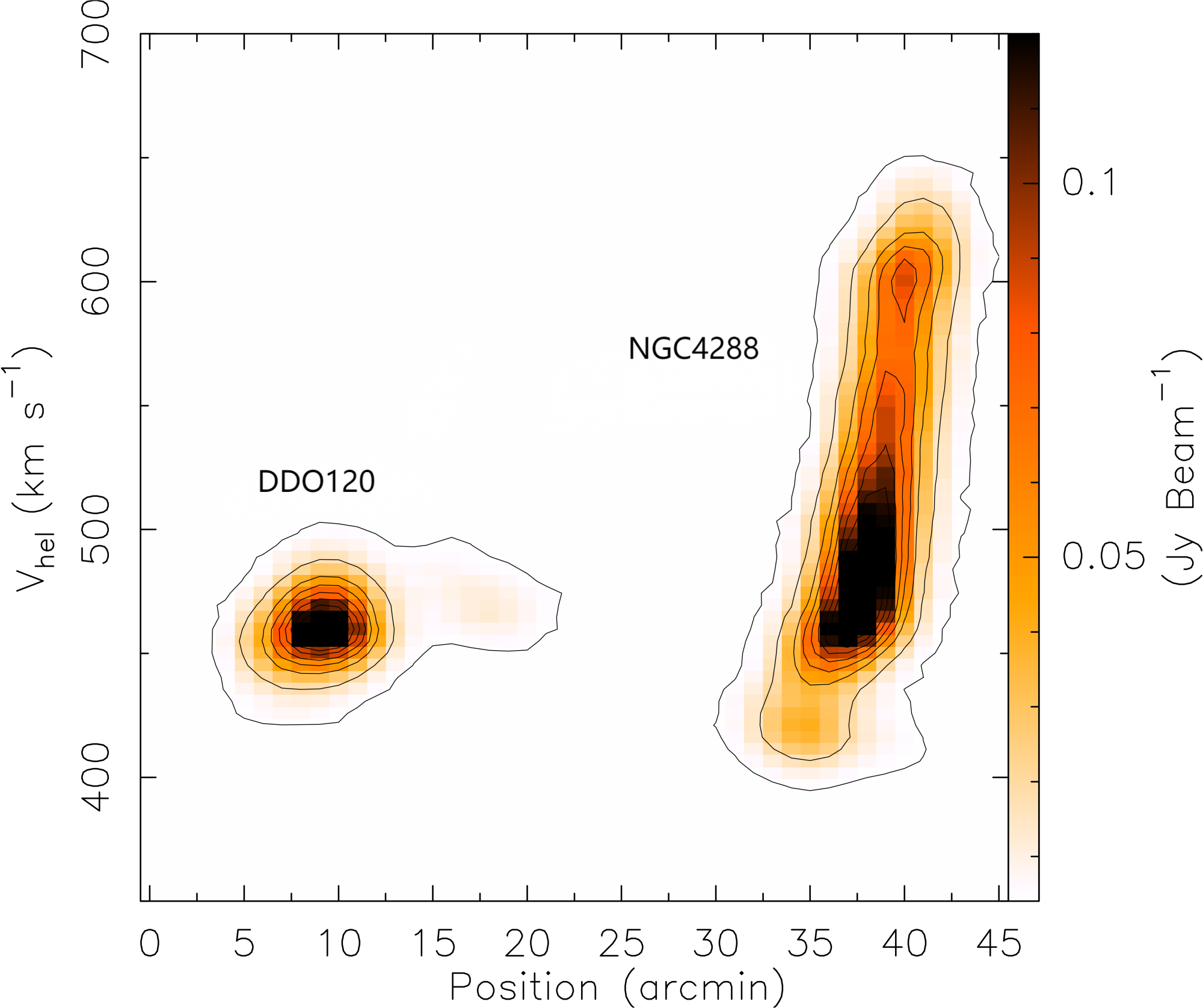}
      \caption{}
    \end{subfigure}%
    \caption{ Top-left: The H I integrated intensity map of M 106 (445 km s$^{-1}$-483 km s$^{-1}$). The unit on the color map is Jy/beam km s$^{-1}$. The contour levels are 0.05, 0.2, 1.5, 10 Jy/beam km s$^{-1}$. Top-right: The FUV and NUV false-color image of NGC 4288  constructed from GALEX data obtained from https://www.legacysurvey.org. Bottom: The PV diagram of DDO 120-NGC 4288. Contour levels start at 0.004 Jy/beam in steps of 0.017 Jy/beam. The direction is shown as a red arrow in the left figure on the top panel. It starts at RA=12:21:28.5, Dec=45:38:43.4 and ends at RA=12:20:38.1, Dec= 46:24:18.9.}
    \label{Fig5}
\end{figure*}

Beside the HI stream connecting NGC 4288 to M 106, we also detect new clouds between DDO 120 and NGC 4288 shown in the top panel of Figure~\ref{Fig5} (Left), which displays the HI integrated intensity map of M 106 for the velocity range from 445 km s$^{-1}$ to 483 km s$^{-1}$.
There are 3 clouds in this region, which seem to be connecting DDO 120 and NGC 4288, as listed below:

(1) One is NGC 4288B, which was first observed by \cite{Wilcots1996AJ} using the Very Large Array (VLA) and confirmed to be associated with NGC 4288 and the HI stream extended toward the northeast of NGC 4288 according to the FAST map in \cite{Zhu2021ApJ}. We measured a total integrated flux of 2.0 Jy km s$^{-1}$ for NGC 4288B, corresponding to an HI mass of $2.7\times 10^{7}M_{\odot }$.

\begin{figure*}
   \centering
\includegraphics[width=0.95\textwidth]{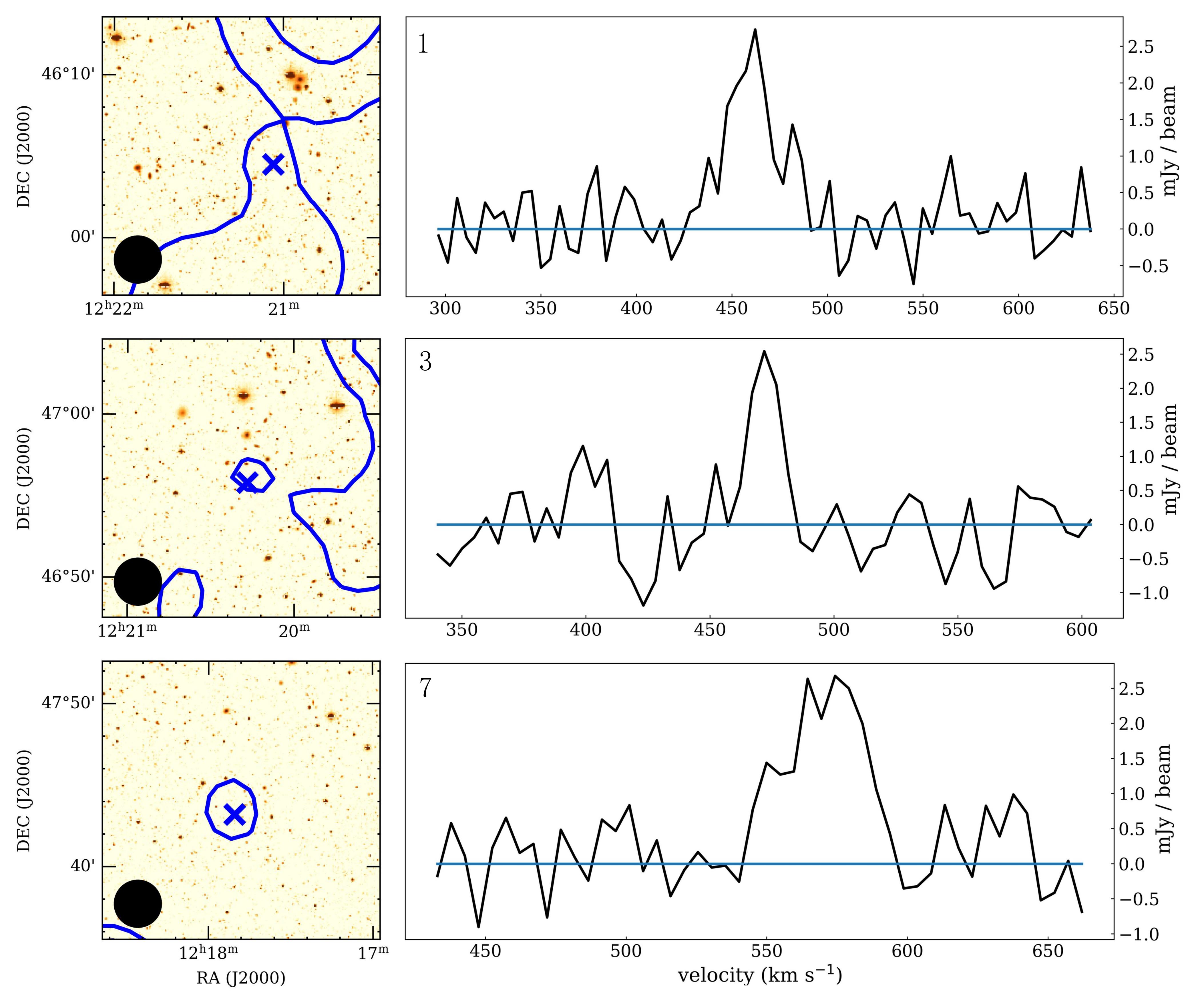}
   \caption{Left: The FAST detected HI integrated flux density contours overlaid on a DESI-LS optical image for Cloud 1 (integrated velocity range is 389-540 km s$^{-1}$, contour level is 0.06 Jy/beam km s$^{-1}$), Cloud 3 (433-506 km s$^{-1}$, contour levels are 0.042 Jy/beam km s$^{-1}$ and 0.091 Jy/beam km s$^{-1}$), and  Cloud 7 (526-613 km s$^{-1}$, contour levels are 0.046 Jy/beam km s$^{-1}$ and 0.091 Jy/beam km s$^{-1}$). Right: Spectra taken at the peak column density.}
  \label{Fig6}
  \end{figure*}

(2) A second HI cloud, marked as Cloud 1, is weak and shown in Figure~\ref{Fig3}, Figure~\ref{Fig5}, and the top panel of Figure~\ref{Fig6}. Figure~\ref{Fig6} presents the spectrum of it which shows a clear detection with 5 sigma. The FAST observation of this cloud resulted in an HI flux of 0.11 Jy km s$^{-1}$.

(3) A third HI cloud, marked as Cloud 2, is located in the northern of DDO 120 as shown in Figure~\ref{Fig5}. It contains two small gas clumps. FAST measured a total integrated flux of 1.4 Jy km s$^{-1}$, or HI mass of $1.9\times 10^{7}M_{\odot }$.

Checking the DELCal image we found no optical counterparts associated with the HI clouds outside DDO 120. The optical image of DDO 120 shows a disturbed disk.
This galaxy is an irregular galaxy with absolute B-band magnitude $M_{B}=-16.50$. The total integrated flux observed by FAST is 8.50 Jy km s$^{-1}$, which is larger than that of the WSRT result (3.69 Jy km s$^{-1}$, \#58, \citealt{KovaC2009}). VLA HI image (Figure 3 in \citealt{Borthakur2011ApJ} ) shows that the HI distribution is not along the optical disk, but perpendicular to the disk. This provides strong evidence that the HI disk is disturbed by the tidal interaction, most likely with NGC 4288 and M 106.  Meanwhile, the GALEX FUV and NUV image of NGC 4288 show tidal features outside the disk of NGC 4288 (See right figure in Figure~\ref{Fig5}(a)). 

\begin{figure*}
    \centering
    \includegraphics[width=0.95\textwidth]{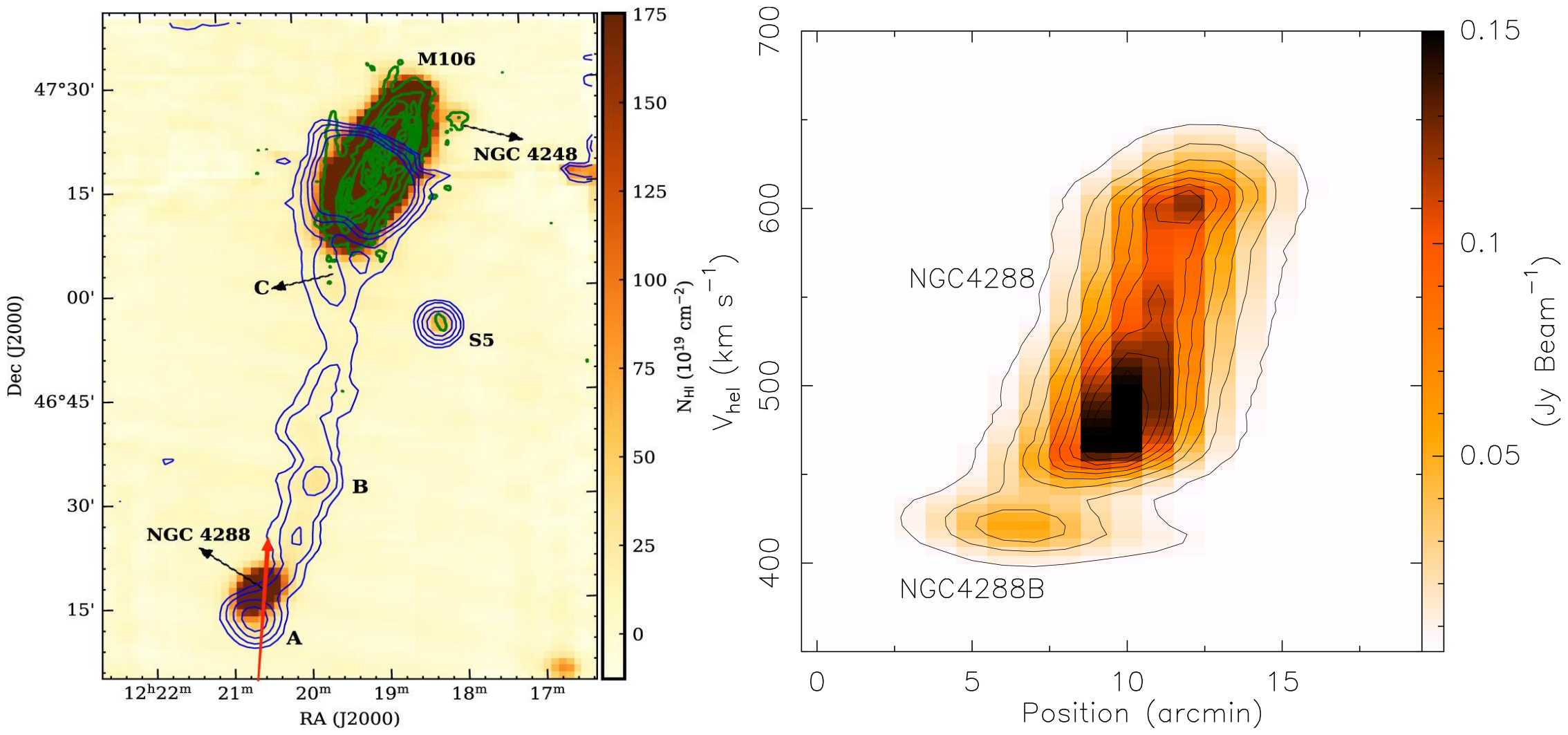}
    \caption{ The position-velocity diagram. Left: The integrated intensity map of M 106 region with the HALOGAS contours overlaid on it (for details see Figure 1 in \citealt{Zhu2021ApJ}). The red arrow marks the direction and position of the PV diagram in the right panel. It starts at RA=12:20:40.6, Dec46:05:56 and ends at RA=12:20:38.7, Dec= 46:26:46. Right: The contours start at 0.004 Jy/beam in steps of 0.017 Jy/beam.}
    \label{Fig7}
\end{figure*}

In Figure~\ref{Fig7}, we have made a position-velocity (PV) diagram cutting through NGC 4288 and NGC 4288B. This figure shows that the HI gas of NGC 4288B appears to be rotating with NGC 4288 disc. In fact, it could be just part of the gas being knocked out from NGC 4288 due to interaction or even direct collision with DDO 120. The P-V diagram of DDO 120-NGC 4288 in Figure~\ref{Fig5}(b) shows a tail-like structure pointing toward NGC 4288. The strong alignment implies that DDO 120 might be undergoing tidal interactions with NGC 4288, or there could be a significant connection between them. 
All these new features discussed above suggest that DDO 120 may actually interact with NGC 4288 to produce the HI stream extending from NGC 4288 to M 106. Such a structure is similar to the LMC-SMC system.
However, compared to the LMC-SMC system, the NGC 4288-DDO 120 system is much further away ($\sim$ 190 kpc projected distance) from M 106 and it may not be bounded by M 106. Also, NGC 4288 and DDO 120 is a loose pair compared with the close pair LMC-SMC. The distance between SMC and LMC is$\sim$ 10 kpc while the projected distance between NGC 4288 and DDO 120 is $\sim$ 60 kpc.
It is interesting to investigate the evolution of the NGC 4288/DDO 120 pair. If these two galaxies can be accreted into M 106, the accretion of HI gas can increase by $8.1\times 10^{8}M_{\odot }$. We will discuss the nature of this system in more detail in section 4.1.

\subsection{Other new HI clouds detected by FAST}

Besides the main HI stream connecting NGC 4288 and M 106, we did not identify any larger diffuse gas structures in this group. There are some extended weak emissions near S 05, but the total HI gas of this region is only $4.3 \times 10^{7} M_{\odot}$. Therefore, the contribution of diffuse gas in this area is limited. Assuming an upper limit of 3 sigma of 1.5 mJy/beam, for an area of 60'x30', which is 181 beams, the upper limit of HI flux that could be missed by FAST would be 1.5 mJy/beam $\times$ 181 beams $\times$ 5 km s$^{-1}$ = 1.36 Jy km s$^{-1}$. Thus the maximum total mass of diffuse gas missed by FAST is unlikely to exceed $1.8 \times 10^{7} M_{\odot}$.

In the M 106 region, to the south of S 05, FAST observed an isolated HI cloud (Cloud 5, RA = 12h17m52.2s, Dec = $46^{\circ}40'42.9"$) with a central velocity of 446 km s$^{-1}$. This gas cloud is clearly visible in  Figure~\ref{Fig2}, Figure~\ref{Fig3}, and Figure~\ref{Fig5}. Checking the DELCal image we found no optical counterparts associated with it.
It is most likely a dark HI cloud or galaxy with very low surface brightness. The HI flux detected by FAST is 0.17 Jy km s$^{-1}$, corresponding to an HI mass of $2.3 \times 10^{6} M_{\odot}$. Cloud 5 is projected at a distance of approximately 15.2' or 33.6 kpc from S 05. It could be formed due to interaction between the tidal tail and S 05.

FAST also detected several other HI filaments or clouds, which we list in Table~\ref{tab1} and mark their positions in Figure~\ref{Fig1} and Figure~\ref{Fig3}. Figure~\ref{Fig6} shows the FAST detected HI integrated flux density contours overlaid on a DESI-LS optical image for Cloud 1, Cloud 3, and Could 7. Their spectra taken at the peak column density are also presented in Figure~\ref{Fig6}. The total mass of these clouds is relatively small, contributing insignificantly to the overall amount of gas that can be accreted into M 106. 

\subsection{Satellite galaxies in the M 106 group}
\subsubsection{FAST observations of the satellite galaxies}

S 14 is the faintest galaxy in our M 106 satellite galaxy catalog, with a B-band magnitude of $M_{B}=-10.8$, located in the southeastern region of M 106. FAST detected an HI flux of 0.085 Jy km s$^{-1}$ for S 14, which corresponds to an HI mass of $1.1\times 10^{6}M_{\odot}$. The HI emission from S 14 is too faint to be shown in the HI integrated flux map Figure~\ref{Fig1}.

NGC 4248 is very close to M 106 and it shows signs of interaction with it. We measured a total integrated HI flux of 3.2 Jy km s$^{-1}$, or HI mass of $4.4\times 10^{7}M_{\odot }$ for NGC 4248, which is slightly more than the HALOGAS measurement (2.7 Jy km s$^{-1}$ or $3.7\times 10^{7}M_{\odot }$ after primary beam attenuation correction, \citealt{Kamphuis2022A&A}).

S 05 is a low surface brightness (LSB, $M_{B}=-13.6$) galaxy located in the southeast of M 106. It has a projected distance of $\sim$ 53 kpc away from M 106. FAST integrated HI flux for S 05 is 3.1 km s$^{-1}$, or HI mass of $4.3\times 10^{7}M_{\odot }$, which is consistent with the HALOGAS measurement (3.1 Jy km s$^{-1}$).

UGC 7356/S 07 is also an LSB ($M_{B}=-13.32$) dwarf galaxy, situated to the south of M 106 and to the east of S 05. The projected distance from M 106 is approximately 28 kpc. The HI flux observed by FAST is 3.45 Jy km s$^{-1}$, corresponding to an HI mass of $4.7\times 10^{7}M_{\odot }$. HALOGAS did not detect this dwarf galaxy. Note that UGC 7356 is close to M 106, and the observed HI flux by FAST might include the clouds at the edge of the M 106 disk.

UGC 7391 is located to the south of M 106 at a projected distance of approximately 187 kpc. This galaxy is situated in the western position between NGC 4288 and DDO 120. Its B-band absolute magnitude is $M_{B}=-15.27$. The HI flux observed by FAST is 2.95 Jy km s$^{-1}$, or HI mass of $4.0\times 10^{7}M_{\odot }$. This galaxy was also detected in the WSRT CVn Survey (\# 57, \citealt{KovaC2009}). The results from FAST and WSRT (2.50 Jy km s$^{-1}$) are in rough agreement.

\begin{figure*}
   \centering
  \includegraphics[width=0.95\textwidth]{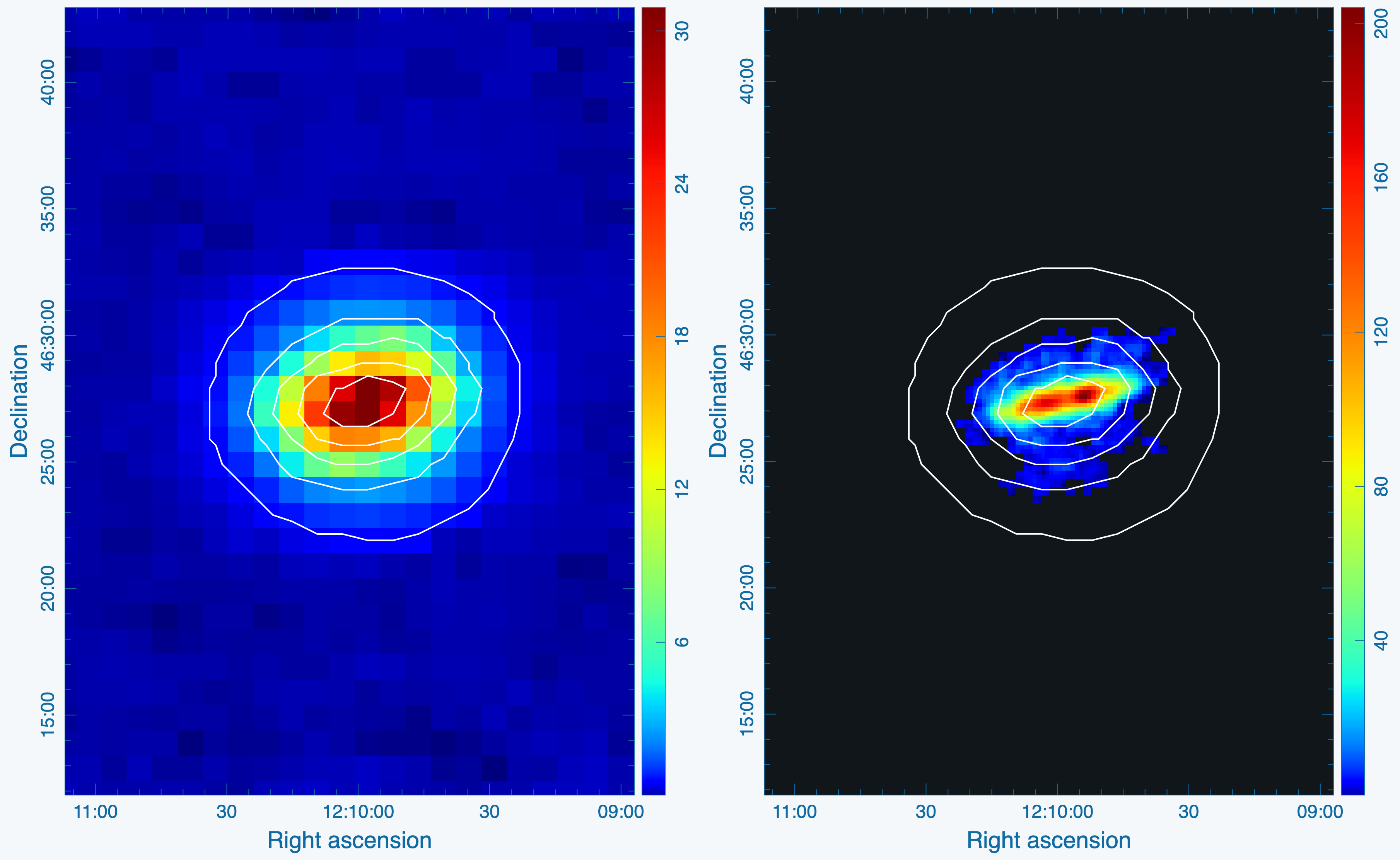}
   \caption{Left: The FAST H I integrated intensity map of NGC 4144 region with  contours overlaid on it. The white contours are integrated over 129-397 km s$^{-1}$ and contour levels are 0.5, 3, 8, 15, 25 Jy/beam km s$^{-1}$. The unit on the color map is Jy/beam  km s$^{-1}$. The rms is 0.025 Jy/beam  km s$^{-1}$.
   Right: The 30 arcsec resolution WSRT interferometer map from
WHISP (\citealt{Van2001ASPC..240..451V}) with the same FAST contours from the left figure. The unit on the color map is w.u. km/s (westerbork unit, 1 w.u. = 5 mJy/beam). 
   }
 \label{Fig8}
\end{figure*}

NGC 4144 is the second brightest member ($M_{B}=-18.45$) among the satellite galaxy candidates and appears to be a normal disc galaxy.
Its optical size is about 5.2'$\times$1.3' with an inclination of 86.2 degrees from the LEDA online database. 
For such an edge-on disk galaxies, the HI contours are expected to exhibit similar distribution, with a large ratio between the major and minor axes.
However, 
Figure~\ref{Fig8}(Left) shows the FAST HI integrated intensity map with its  contours overlaid on it for NGC 4144 (integrated velocity range from 129 km s$^{-1}$ to 397 km s$^{-1}$), which does not meet our expectation. Figure~\ref{Fig8}(Right) shows a deep WSRT imaging from WHISP (\citealt{Van2001ASPC..240..451V}) overlaid with FAST contours from the left figure. It shows the detailed HI distribution of the galaxy disk with $30'' $ resolution.
The FAST map shows NGC 4144 has an HI disk of about $11.0'$ long and $6.0'$ wide, and the extended HI gas above and below the plane of the galaxy's disk is
much more extended than that seen in the WSRT map.
From Figure~\ref{Fig8}, it can be observed that the contour lines of NGC 4144 
are much more extended in the vertical direction for unknown reasons. 
\cite{Schechtman} modeled NGC 4144 and suggested a disk plus bulge model. Although their model increases the height of the galactic disk (550 pc), roughly twice that estimated from other studies (e.g., \citealt{Seth2005AJ}), it is not sufficient to explain the vertical extended profiles of the HI distribution. The SFR of NGC 4144 is 0.17 $M_{\odot }$ yr$^{-1}$ (\citealt{KAR13SFR}), which is lower than the predicted SFR for a main sequence galaxy with comparable stellar mass (0.26 $M_{\odot }$ yr$^{-1}$, according to \citealt{Renzini2015ApJ}). Therefore, it is unlikely that the 
extraplanar gas
is caused by gas outflows from star formation. It is more likely influenced by the interactions or gravitational distortion from neighboring galaxies within the group, which move more disk gas to the halo. In conclusion, the unusual vertical extension of NGC 4144's HI distribution could be considered as evidence for it to be a satellite galaxy of M 106. The total HI integrated flux of NGC 4144 measured by FAST is 59 Jy km s$^{-1}$, corresponding to an HI mass of $7.9\times 10^{8}M_{\odot }$.

NGC 4242 is also an extended source. It is a bright member($M_{B}=-17.46$) in the satellite galaxy candidates. It has an HI disk of about 8.2 $\times$ 5.2' (\#56, \citealt{KovaC2009}). Its optical size is about 3.8'$\times$2.7'. The FAST detected HI flux  is 42 Jy km s$^{-1}$, with an H I mass of $5.6\times 10^{8}M_{\odot }$. UGC 7401 is also detected by both FAST and HALOGAS. However, the SBF distance for UGC 7401 is larger than 12.7 Mpc and confirmed to be a background galaxy by \cite{Carlsten2021ApJ}. Thus this galaxy is not included in our analysis.

\subsubsection{The properties of satellite galaxies}
With the above FAST HI data for satellite candidates in M 106, we can study the HI properties of satellite galaxies.
The top panel of Figure~\ref{Fig10} shows the distribution of 
specific SFR (sSFR, defined as SFR/$M_{Star}$ )
against the stellar mass $M_{Star}$ in log units for the satellite galaxies in M 106. The open circles represent sSFR based on $H\alpha$ flux and filled circles represent sSFR based on FUV flux, which are color-coded by their projected distances from M 106. Except for DDO 120, sSFR based on $H\alpha$ and FUV values are generally in good agreement with each other for satellite galaxies with duplicated measurements. We plotted the SFMS from \cite{Renzini2015ApJ} with a scatter of 0.2 dex displayed as the shaded region for comparison,
which calculated galaxies' SFRs based on their stellar mass using a galaxy sample selected from Sloan Digital Sky Survey (SDSS) DR7 release (\citealt{Abazajian_2009}) with reliable spectroscopic redshift lying at 0.02-0.085:
\begin{equation}
 log(SFR)=(0.76\pm 0.01)log(M_{Star}/M_{\odot }  )-7.64\pm 0.02   \tag{3}
\end{equation}

We also use the formula from \cite{Guo_2021} to classify the star-forming galaxies and quenched galaxies as follows (shown as the red line):
\begin{equation}
log(SFR_{cut})=0.65log(M_{Star}/M_{\odot }  )-7.25   \tag{4}
\end{equation}
It is clear that bright satellite galaxies($M_{B}<-10.0$) in M 106 show a significant deviation from SFMS. At the high-mass end, satellite galaxies lie closer to the SFMS than at the low-mass end. 
Furthermore, bright satellite galaxies in the M 106 group with a larger projected distance tend to exhibit higher SFR. 

The bottom panel of Figure~\ref{Fig10} shows the relationship between the HI mass of satellite galaxies and the diameter of the galaxy converted to kpc in the
B-band at the 25.0 mag arcsec$^{-2}$ ($D_{25}$). The dashed grey line with a scatter of 0.23 dex represents the linear fit line for the field galaxies sample including all types of morphologies (Table 5, \citealt{Haynes1984AJ}). Our luminosity-cut sample generally follows the average $M_{HI}$-$D_{25}$ relation observed in field galaxies, displaying some scattering around it, which indicates that bright satellites of M 106 have not experienced much physical mechanisms that can remove its cold gas. 

\subsubsection{Selection effects}
Our analysis of HI gas properties in the group satellites has been limited to galaxies brighter than $M_{B}=-10.0$, which could bias the result in favor of bright galaxies. Dwarf satellites fainter than $M_{B}>-10.0$ (shown in Table~\ref{tableA1}) are more vulnerable to external influences and may provide a more sensitive diagnostic, and thus should also be considered in our investigation. However, after searching the literature and online database, we found that there is a significant lack of data on the parameters of the faint galaxies ($M_{B}>-10.0$) within the M 106 group, including parameters such as stellar mass, $logD_{25}$, and SFRs. Additionally, the membership status of some of these galaxies remains highly uncertain. This scarcity of information presents challenges in incorporating these smaller galaxies into studies of HI properties within the M 106 group. 

The 16 faint satellite galaxy candidates are not detected by FAST. Assuming an upper limit of 3$\sigma$ of 1.5 mJy level for an HI undetected satellite galaxy with 100 km s$^{-1}$ velocity width, the maximum total HI mass for it is $2.0\times 10^{6}M_{\odot }$.
Using the r-band luminosity as a proxy for stellar mass, we can use the $M_{HI}/L_{r}$ ratio to indicate the HI richness in a galaxy. For the faint galaxies in the M 106 group, the average $log(M_{HI}/L_{r})$ is -0.64, with a median value of -0.8. In contrast, the average $log(M_{HI}/L_{r})$ for the sample of 11 bright galaxies is 0.13, with a median value of 0.33. These findings suggest that the M 106 group may contain a significant number of HI-poor dwarf satellite galaxies,  in contrast to the bright satellites sample, which are HI-normal comparing to field galaxies.
Unfortunately, the lack of information on stellar mass, $logD_{25}$, and SFRs for the faint galaxies in M 106 group makes it difficult to draw solid conclusions regarding the deviation  from the SFMS and the $M_{HI}$-$D_{25}$ relation observed in field galaxies. Therefore, it is important to note that our findings regarding the HI properties of satellites in the M 106 group are restricted to the bright galaxy members.


\section{Discussion}
\subsection{Formation of the M 106 HI stream, an analog of the Magellanic stream?}
The morphology and HI mass of the HI stream between NCG 4288 and M 106 is similar to that of the Magellanic stream. The major difference is that NGC4288 and DDO 120 are much further away from M 106, and it is not known if these two galaxies are bound to each other. For NGC 4288, the line width $W_{50}$ is 174 km km s$^{-1}$ and $W_{20}$ is 241 km s$^{-1}$ (\citealt{Springob2005}). Based on the method described in \cite{Guo2020}, we estimate that its dynamical mass is about $5.5\times 10^{10}M_{\odot }$. Note that we only know the radial velocity(one-dimensional velocity) of DDO 120 in the NGC 4288-DDO 120 system. We assume their possible velocity distribution is isotropic, then the three-dimensional velocity of DDO 120 in this system is $V_{rot}=\sqrt{3}\left | V_{1}-V_{2} \right |$  =124.7 km s$^{-1}$.
Thus a satellite located within  $ r=2GM/V_{rot}^{2} $=30.5 kpc away from NGC 4288 is considered to be bounded. However, DDO 120 is 55 kpc away in projection, which means DDO 120 can not be bound to NGC 4288 under isotropic conditions. We introduce an isotropic parameter a($\ge$ 1) and b($\ge$ 1) for the relative velocity and distance for DDO 120(e.g., a=1 means DDO 120 moves in the line of sight direction. $a^{2}=3$ means an isotropic condition). Then the corresponding three-dimensional velocity and distance of DDO 120 in the system is 72a km s$^{-1}$ and 55b kpc respectively. The physical parameter $V_{rot}^{2} r< 2GM$ would represent a bounding condition, which leads to $ba^{2}< 1.64$. The bounding condition can be met in certain situations (e.g., a=1, b=1). Although NGC 4288 only has a TF distance of 8.0 Mpc with more than 20$\%$ uncertainty (\citealt{Tully2008ApJ}), the features of the HI steam reported by \cite{Zhu2021ApJ} strongly suggests a connection between the HI steam and M 106/NGC 4288. It is very likely that NGC 4288 is at the same distance as M 106. DDO 120 has a distance of $7.73\pm 0.05$  Mpc measured by the TRGB method (\citealt{Tikhonov2018AstBu}), which is virtually at the same distance as M 106. Thus M 106, NGC 4288, and DDO 120 are probably at the same distance. To meet the parameter conditions $ba^{2}< 1.64$, which is equivalent to a<1.28 when DDO 120 and NGC 4288 are at the same distance, the relative velocity of DDO 120 in the NGC 4288-DDO 120 system should mainly concentrated in the radial direction. Under this condition, the orbit of DDO 120 around NGC 4288 is perpendicular to the plane facing us. As mentioned in section 3.3, the HI disk of DDO 120 is perpendicular to its optical disk. This could happen if DDO 120 moves around NGC 4288 and the HI disk gets disturbed due to tidal interaction. Meanwhile, the 3D view of the accretion stream shown in Figure 2 in \cite{Zhu2021ApJ} appears as a vertical bar. Thus the NGC 4288-DDO 120 system probably moves in the line of sight direction around M 106 if the HI stream originates from the NGC 4288-DDO 120 system. 

\begin{figure*}
   \centering
   \includegraphics[width=0.8\textwidth] {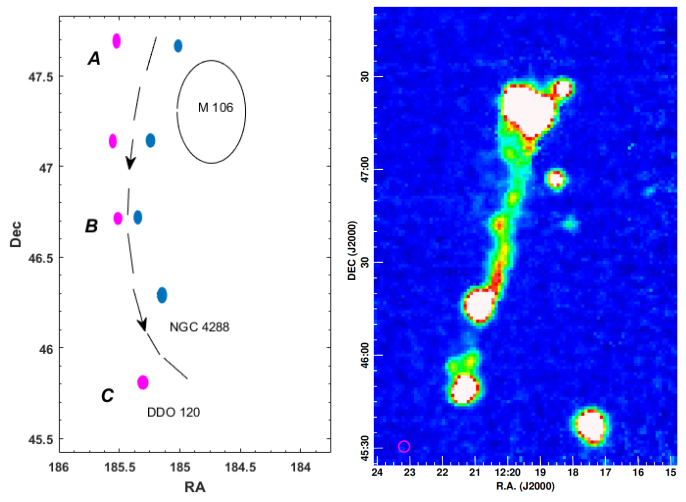}
   \caption{Left: The possible projected trajectories of NGC 4288 and DDO 120 within the M 106 galaxy group. The circles represent the optical disk diameters of each galaxy. NGC 4288 is shown in blue, and DDO 120 is shown in purple. Right: FAST HI intensity map. The velocity integration range is 389-496 km s$^{-1}$.
   }
   \label{Fig9}
\end{figure*}

Based on the above analysis, we proposed an evolution scenario for NGC 4288-DDO 120. The movement of this system is perpendicular to the M 106-NGC 4288-DDO 120 plane facing us. Figure~\ref{Fig9} shows a projected view of the movement. In an early stage, DDO 120 is attracted by NGC 4288 when these two galaxies are captured by M 106. When the system NGC 4288-DDO 120 moves from position A to B, these two galaxies get closer and closer. And finally, in some places near position B, two galaxies strongly interact with each other, and direct collisions are also possible. The atomic gas in NGC 4288 and DDO 120 gets extracted from the galaxies due to tidal forces and gets distributed in the intra-group medium. This scenario explains why most of the filamentary gas are found near position B. By the time NGC 4288-DDO 120 moves to the present-day position C, the atomic gas extracted out of the satellite galaxies are attracted to the gravitational potential center M 106 and formed the HI stream connecting NGC 4288 and M 106. The time scale for this scenario is about 130 kpc/100 km s$^{-1}$=1.3 Gyr. Although further detailed numerical analysis is required to determine the origin of the HI filament, the HI features in this region provide support for the hypothesis that NGC 4288 and DDO 120 are analogs of the Magellanic clouds.

\subsection{Quenching mechanisms in M 106 group}
\begin{figure*}
    \centering
    \begin{subfigure}[b]{0.95\textwidth}
      \includegraphics[width=\textwidth]{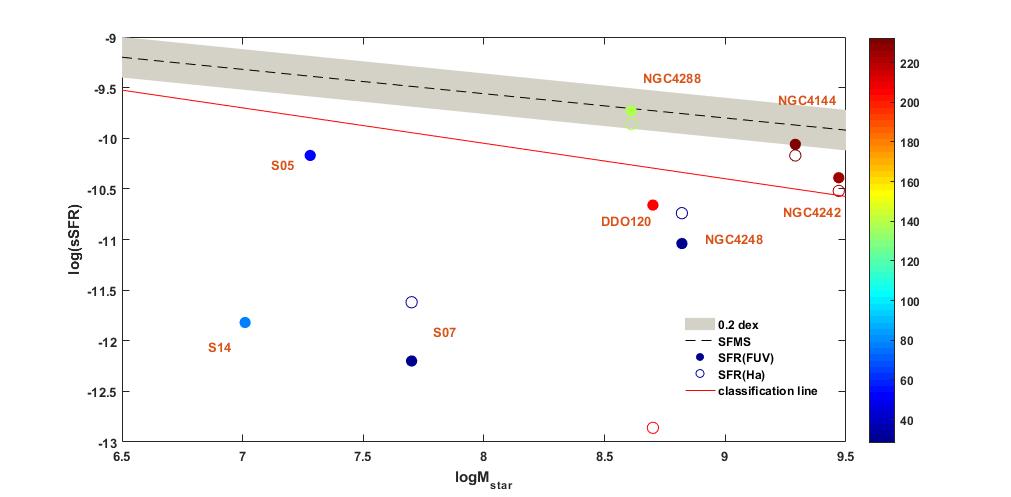}
      \caption{}
    \end{subfigure}%
    \\
    \begin{subfigure}[b]{0.95\textwidth}
      \includegraphics[width=\textwidth]{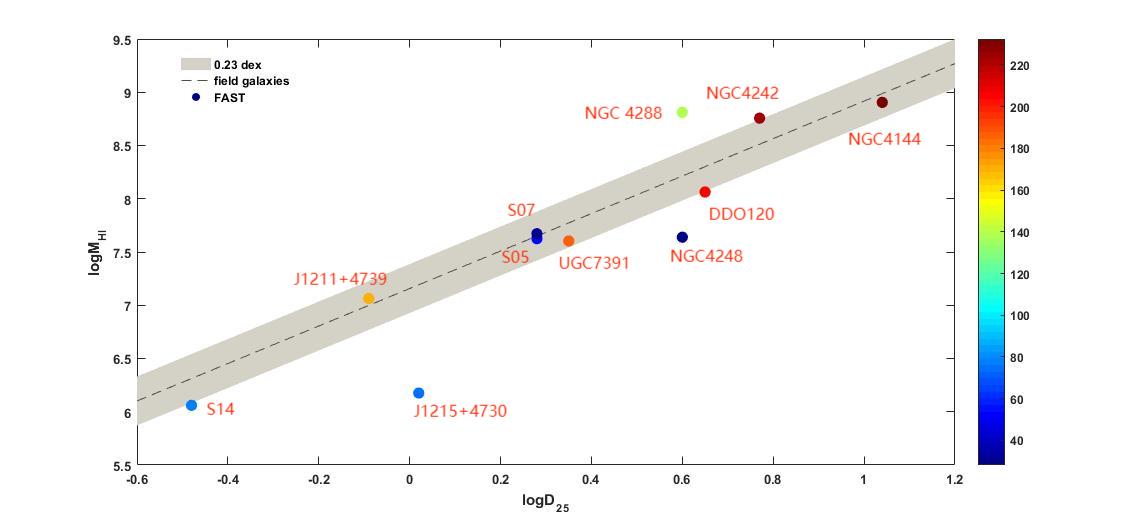}
      \caption{}
    \end{subfigure}%
    \caption{The sSFR and HI mass of satellite galaxies in M 106. Top: sSFR vs. stellar mass $M_{Star}$, color-coded by their projected distances from M 106. The sSFR based on the observed FUV or $H\alpha$ flux extracted from \protect\cite{KAR13SFR} is plotted as filled or open circles separately. As a reference, the fitted star-forming main sequence relation from \protect\cite{Renzini2015ApJ} is given as a dashed gray line, with a scatter of 0.2 dex given as shading. The red line separates the star-forming and quenched
    galaxies. Bottom: $M_{HI}$ vs. $D_{25}$ color-coded by their projected distances from M 106. HI detections by FAST are plotted as filled circles. The dashed grey line represents the linear fit line for the field galaxies sample in \protect\cite{Haynes1984AJ}. Its scatter of 0.23 dex is given as shading.}
    \label{Fig10}
\end{figure*}

Our sample of satellite galaxies in the M 106 group spans a stellar mass range from $10^{7.01}M_{\odot}$ to $10^{9.47}M_{\odot}$. 
For dwarf galaxies below a stellar mass threshold of $M_{Star} \lesssim10^{9} M_{\odot }$, quenched galaxies are rarely found in the field (\citealt{Haines2007MNRAS,Haines2008};\citealt{Peng2010ApJ};\citealt{Geha2012ApJ};\citealt{Wetzel2012MNRAS};\citealt{Woo2013MNRAS}). 
Thus the quenching scenario is expected to occur after these galaxies have joined the group. 
Environmental quenching mechanisms including ram pressure stripping(RPS;\citealt{Gunn1972ApJ};\citealt{Rasmussen2008MNRAS}), tidal forces(\citealt{Haynes1984AJ};\citealt{Freeland2009AJ}), starvation(\citealt{Larson1980ApJ}), and galaxy harassment(\citealt{Barazza2002}), can effectively suppress the SFRs in satellite galaxies by reducing the cold gas content which is the fuel for star formation. 
However, satellite galaxies in M 106 have cold gas content similar to field galaxies (shown in  Figure~\ref{Fig10}b). A direct speculation is that these satellites have acquired new gases during their subsequent evolution after quenching, or the quenching mechanism is not dominated by removing or depleting the gas.

Recent studies of galaxies in the local Universe have given us a general picture of the timescales associated with the dominant quenching mechanisms affecting satellite galaxies across various mass ranges.
For the least massive ('ultra-faint') dwarf satellites with stellar masses around $10^{5}M_{\odot }$ in  Milky Way-like and more massive host haloes($> 10^{12} M_{\odot }$), the dominant quenching mechanism is reionization(e.g.,\citealt{Bullock2000ApJ};\citealt{Weisz_2014};\citealt{Rodriguez2019MNRAS};\citealt{Sand2022ApJ}). 
For the intermediate-mass dwarf satellites ($10^{5} M_{\odot } \le M_{Star} \le  10^{8} M_{\odot } $), RPS or tidal stripping can directly strip their interstellar media with a short quenching time scale ($\sim$ 2Gyr). 
However, for galaxies with stellar mass above $10^{8}M_{\odot }$, satellites exhibit considerable resistance to stripping forces, leading to quenching occurring over extended time scales($\sim$ 4-6 Gyr;\citealt{Wetzel2012MNRAS};\citealt{Wheeler2014MNRAS}). And this time scale is in accordance with the starvation mechanism(\citealt{Larson1980ApJ};\citealt{Fillingham2015MNRAS}). 

S 05, S 07, and S 14 have intermediate stellar mass($10^{7.01} M_{\odot } \le M_{Star} \le  10^{7.7} M_{\odot } $) and lie in close distance to M 106 and the HI stream(see Figure~\ref{Fig1} and Figure~\ref{Fig3}). These galaxies have quenched sSFRs (Fig~\ref{Fig10}a), but they are not gas-poor (see Fig~\ref{Fig10}b).
The excess gas of S 07 after quenching may be attributed to the source confusion, as discussed in section 3.4.1, or to gas accretion from the cloud at the edge of the M 106 disk.
On the other hand, the velocity continuity observed in the region between S 05, S 14, and the HI stream, coupled with the presence of an isolated HI cloud(Cloud 5), implies the potential existence of additional low-column density gas below the observational limit of FAST. 
A straightforward speculation is that S 05 and S 14 are accreting gas from the HI stream.
In this case, we might just capture the intermediate phase between the quenching of the galaxies and its reactivated star formation due to the accretion of gas.
Future study of the star-formation history of these two galaxies and deeper HI map in this region are needed to confirm or reject this scenario.

NGC 4248 is gas-poor and quenched. It has a relatively large stellar mass ($10^{8.82}M_{\odot }$) and a projected distance of $\sim$ 29 kpc away from M 106. FAST HI map shows signs of interaction between NGC 4248 and M 106. Thus a combination of quenching mechanisms including RPS, tidal forces, and starvation is likely to have played a role.

NGC 4288 is a gas-rich galaxy and actively forming stars with a stellar mass of $10^{8.61}M_{\odot }$.
According to our discussion in section 3.2 and section 4.1, it is accreting gas from DDO 120, which has relatively low HI gas and quenched SFR. The stellar mass of DDO 120 is $10^{8.7}M_{\odot }$. This galaxy pair is probably on its first passage around M 106. The time scale is $\sim$ 1.3 Gyr, shorter than the time scale for starvation. Furthermore, 
the distance between these two galaxies and M 106 is larger than 140 kpc. This distance scale exceeds the maximum distance($\sim$ 90 kpc) at which ram pressure can efficiently strip gas from dwarf galaxies in Milky Way-like galaxies(\citealt{Gatto2013MNRAS}).
Thus DDO 120 is probably quenched by interactions with NGC 4288.

NGC 4144 and NGC 4242 have stellar masses larger than  $10^{9}M_{\odot }$ and are located nearly at the edge of the group virial radius. Considering that they are rich in HI gas, these two galaxies are probably recent additions to the M 106 group, 
just beginning to be affected by the starvation mechanism, so the sSFR is relatively low, but it has not reached the level of being quenched yet.

Nevertheless, besides environmental quenching mechanisms, other mechanisms may also lead to the suppression of star-forming activities in NGC 4144 and NGC 4242. 
One possibility is tidally induced bars,
which typically form in the disk of the dwarf galaxy during its first encounter with the host(\citealt{Klimentowski2009MNRAS}). 
Bars can induce shocks and shear. These effects can stabilize the gas against collapse by increasing turbulence, and consequently quench the star formation activity in the galaxy(e.g.,\citealt{Tubbs1982ApJ};\citealt{Haywood2016A&A};\citealt{Khoperskov2018A&A}).
The time scale for bar quenching in disk dwarf galaxies orbiting in the static potential of a larger galaxy like the MW
is about 2 Gyr, similar to RPS and tidal stripping(\citealt{Gajda2018ApJ}).

According to the LEDA database, NGC 4144 and NGC 4242 are classified as barred galaxies. If these two galaxies' relatively low sSFRs are  due to bar quenching, this would alleviate any need for the group environment to remove gas from these galaxies.
NGC 4248 is also a barred galaxy that may have experienced bar quenching. Indeed, bar quenching may be present in other satellite galaxies (e.g. S 05, S 07, S 14) as well, but deriving a general statement proves challenging. This uncertainty stems from the difficulty in determining whether these galaxies possess bars, given their low luminosity or their orientation along our line of sight.

Our discussion of the quenching mechanisms in satellite galaxies is primarily based on the HI mass, the stellar mass, the distance from the host galaxy, and the distribution of HI gas in the galaxy group. A more in-depth investigation into quenching mechanisms requires studies on star formation history and dynamical evolution simulations. Meanwhile, other quenching mechanisms such as star formation and stellar feedback from supernovae, should also be considered.

\section{Conclusions}
Our main conclusions are summarized as follows:

(1) We have investigated all existing satellite candidates for M 106 from the literature and online database to get a total list of 64 potential satellites. After removing background galaxies and applying a minimum luminosity cut, we  we obtain a catalog containing 11 potential satellites for M 106. We further categorize these candidates into different classes based on their projected distance and reliability of distance measurements. All the satellite galaxies with M$_B< -10.0$ have HI detection by FAST. 

(2) We present the results of the deepest HI imaging of the entire M 106 galaxy group region achieved to date using FAST. Overall, FAST detects more HI gas than previous studies (18$\%$-130$\%$ more for NGC 7391 and DDO 120, respectively). Compared to HALOGAS, FAST also reveals more HI features around M 106. We summarize the filamentary gas structures, calculate the HI mass of these HI filaments or clouds.

(3) We conducted a comprehensive analysis of the origin of the HI stream reported in \cite{Zhu2021ApJ}. The GALEX FUV+NUV image of NGC 4288 reveals tidal structures, and the VLA HI column density contours for DDO 120 also show HI distribution perpendicular to the optical disk, suggesting that both NGC 4288 and DDO 120 are influenced by tidal forces. FAST deep mapping of this region discovered new HI gas clouds, including 'Cloud 2' extending from the northern end of DDO 120, 'Cloud 1' between DDO 120 and NGC 4288, and NGC 4288B. The properties of these clouds indicate that DDO 120 and NGC 4288 are in interaction, resulting in the HI stream extending from the northern end of NGC 4288 to M 106. Consequently, the total length of the HI stream increases from 160 kpc to approximately 190 kpc, connecting DDO 120, NGC 4288, and M 106, forming a system similar to the Magellanic Stream (SMC-LMC). Considering the vertical distribution of the HI stream, we propose that DDO 120 and NGC 4288 are moving along the line of sight direction around M 106, while the interactions or even collisions between DDO 120 and NGC 4288 produce the gas that we now observe in the HI stream, whose shape is also influenced by the gravity of M 106. After calculating the relevant physical parameters, we proposed a possible evolution scenario for this pair of galaxies.


(4) Focusing exclusively on the bright members($M_{B}<-10.0$), we found that many satellite galaxies \textbf{within M 106 have lower SFR comparing to the SFMS.} Bright satellites at the low-mass end and with closer projected distances to M 106 are more easily quenched. Furthermore, the HI content of satellite galaxies in M 106 exhibits a trend similar to that observed in field galaxies, suggesting that bright satellites within M 106 are normal in HI content.
Based on the new data from FAST, we discuss the possible mechanisms leading to the quenching in the M 106 group. 

\section*{Acknowledgements}
We thank the FAST staff for help with the FAST observations and data reduction.
We thank the anonymous referee for insightful comments
and constructive suggestions. Y.L. is grateful to Yanbin Yang for useful discussions.
This work is supported by
the National Key R$\&$D Program of China (2018YFE0202900;
2017YFA0402600). We also thanks the National Natural Science Foundation of
China (No. 12041301 and No. U1531246).
This research made use of data from WSRT HALOGAS-DR1 and WSRT WHISP. 
The Westerbork Synthesis Radio
Telescope is operated by ASTRON (Netherlands Institute for
Radio Astronomy) with support from the Netherlands Foundation for Scientific Research NWO. 
The Five-hundred-meter Aperture Spherical radio Telescope (FAST) is a Chinese national mega-science facility, funded by the National Development and Reform Commission. FAST is operated and managed by the National Astronomical Observatories, Chinese Academy of Sciences.
\section*{Data Availability}

The raw data used in the article will be published on the FAST website: \href{https://fast.bao.ac.cn}{https://fast.bao.ac.cn}.
The PIDs are N2021$\_$4 and N2022$\_$1.
Please contact the author (mz@nao.cas.cn) for processed data.







\appendix
\section{Notes on individual satellite galaxy candidates}
\begin{figure*}
   \centering
\includegraphics[width=0.97\textwidth]{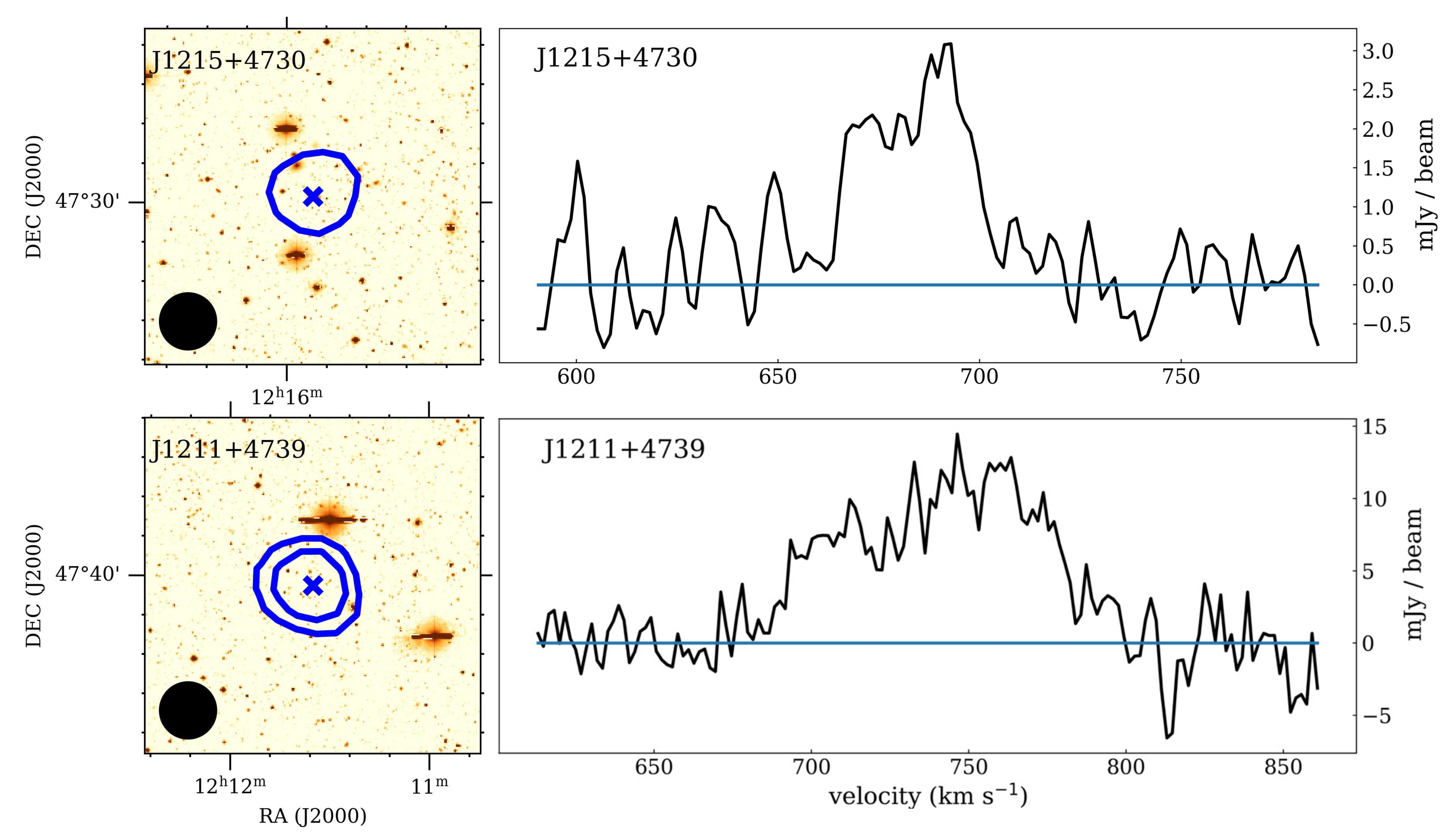}
   \caption{Left: The FAST detected HI integrated flux density contours overlaid on a DESI-LS optical image for satellite galaxies J1215+4730 (integrated velocity range is 641-729 km s$^{-1}$, contour level is 0.07 Jy/beam km s$^{-1}$), and J1211+4739 (664-811 km s$^{-1}$, contour levels are 0.176 Jy/beam km s$^{-1}$ and 0.352 Jy/beam km s$^{-1}$) are shown in the left figure. Right: Spectra taken at the peak column density are shown in the right figure.}
  \label{FigB2}
  \end{figure*}

SDSS J121551.55+473016.8 (J1215+4730) is rejected by \cite{Carlsten2020ApJ} due to the inappropriateness in visual inspection rather than the distance. Our HI observation by FAST shows there exists a faint HI source there. The HI line flux is 0.11 Jy km s$^{-1}$, or HI mass of $1.2\times 10^{6}M_{\odot }$, with a central velocity of 685 km s$^{-1}$(shown in top plane of Figure~\ref{FigB2}). Thus J1215+4730 could possibly be a satellite galaxy of M 106. Another two galaxies dw1219+4718 and dw1219+4727 were confirmed to be satellite galaxies by \cite{Spencer2014ApJ} based on the TRGB distances from \cite{Munshi2007}. However, \cite{Carlsten2021ApJ} questioned the reliability of these TRGB distances and considered dw1219+4718 (>10.1 Mpc) and dw1219+4727 (>12.7 Mpc) to be background galaxies based on SBF results. Meanwhile, the velocity of dw1219+4718 and dw1219+4727 measured by HALOGAS is 811.3 km s$^{-1}$ and 941km s$^{-1}$respectively. The radial velocity of dw1219+4727 far exceeds the velocity range of the M 106 disk, thus it should not be a satellite galaxy of M 106. \cite{Kamphuis2022A&A} classified dw1219+4718 as a companion of M 106 according to its Kitt Peak National Observatory (KPNO) R band image with measured HI line flux 0.06 Jy km s$^{-1}$. FAST also detected an HI source (0.045 Jy km s$^{-1}$, or $6.1\times 10^{5}M_{\odot }$) here, with a central velocity of 814 km s$^{-1}$. As a result, we add  J1215+4730 and dw1219+4718 back to our subsample as possible satellite galaxy candidates. 

\section{The classification of satellites}
We adopt the following classification scheme:

(1) NGC 4248, S 05, and UGC 7356 lie within 100 kpc from the host, with distances measured by different methods(TRGB, TF, and SBF). These  measurements are generally in good agreement with each other (within errors). They are also confirmed to be satellites by \cite{Spencer2014ApJ} that their line-of-sight velocity relative to M 106 falls within the $\pm V_{esc}/\sqrt{3}$ boundaries(see Figure 4 in \citealt{Spencer2014ApJ}), where $V_{esc}$ is the predicted escape velocity. \cite{Spencer2014ApJ} calculated the escape velocity as a function of radius for a Navarro-Frenk-White mass distribution (NFW, \citealt{Navarro1996ApJ}) plus a disk component. Assuming velocity isotropic, they plot the NFW $\pm V_{esc}/\sqrt{3}$ as a function of projected radius in Figure 4.
Thus we suggest NGC 4248, S 05, and UGC 7356 are the most probable satellites among these 11 potential satellites for M 106. We marked them as code 1 in Table~\ref{tableA2}. 

(2) NGC 4288 and S 14 have projected distances within 200 kpc. NGC 4288 only has TF distance ranges from 8 Mpc to 26.10 Mpc (NED) and its line-of-sight velocity with respect to M 106 is within the $\pm V_{esc}/\sqrt{3}$  boundaries. S 14 only has SBF distance (7.9 Mpc) from \cite{Carlsten2021ApJ}. These two galaxies lack corroborating distance verification from either other methods or identical methods, albeit their projected radius are modest. Thus we suggest that NGC 4288 and S 14 are probable satellites. They are marked as code 2. 

(3) NGC 4144, NGC 4242 and DDO 120 have projected radius larger than 200 kpc, but less than 300 Kpc. They have multiple distance measurements with TF and TRGB. The TRGB distances are generally in consistent with TF distances to NGC 4144 and DDO 120. The TF distance to NGC 4242 ranges from 5.20 to 10.40 Mpc (mean value $6.3\pm 1.50$ Mpc), while only one recent TRGB distance to NGC 4242 is $5.3\pm0.3$ Mpc (\citealt{Sabbi2018ApJS..235...23S}). Although we use this TRBG value to represent the distance to NGC 4242, we categorize NGC 4242 together with NGC 4144 and DDO 120 as possible satellites, which are marked as code 3. 

(4) UGC 7391, J1215+4730, and SDSS J121134.99+473927.1 (J1211+4739) have projected distances within 200 kpc. We found no distance information for them. \cite{Spencer2014ApJ} suggest that UGC 7391 and J1211+4739 are possible satellites and J1215+4730 are probable satellites base on their line-of-sight velocity with respect to M 106. 
We marked these galaxies as Code 4 and assigned them a distance of 7.60 Mpc, which corresponds to the distance to M 106 as determined by water maser measurements (\citealt{Humphreys_2013}). This distance is used for objects tagged with "mem", denoting galaxy membership in known groups with measured distances to other member galaxies.
\section{Tables of proprieties for satellites in M 106 }
\section{Figures of spectra}
\begin{table*}  
\centering
\caption{Proprieties for 16 candidate satellite galaxies with low luminosity in M 106}
\label{tableA1}
\resizebox{0.8\textwidth}{!}{%
\begin{tabular}{ccccccc}
\hline
\hline
(1)              & (2)     & (3)     & (4)      & (5)      & (6)                & (7)     \\
Name             & Car2021 & Spe2014 & RA       & Dec      & $D_{SBF}$          & r       \\
                 &         &         & (J2000)  & (J2000)  & Mpc                &         \\ \hline
dw1219+4718      & N       & Y       & 12:19:27 & 47:18:45 & \textgreater{}10.1 & 18.32   \\
dwC              &         &         & 12:10:27 & 46:44:49 &                    & 18.42   \\
588298663037239998$^{a}$&     & X       & 12:25:06 & 47:18:51 &                    & 18.44   \\
LV J1215+4732    &         &         & 12:15:51 & 47:32:56 & 11.7               & 18.67   \\
{[}KKH2011{]}S16 & C       & X       & 12:23:46 & 47:39:33 & 7.4                & 19.2    \\
dw1220+4922      & M       &         & 12:20:14 & 49:22:52 & 6.7                & 19.51   \\
dw1218+4623      & M       &         & 12:18:03 & 46:23:05 & 5.4                & 19.56   \\
dw1223+4848      & M       &         & 12:23:13 & 48:48:56 & 10                 & 20.35   \\
{[}KKH2011{]}S06 & C       & X       & 12:19:06 & 47:43:49 & 7.6                & 20.4    \\
{[}KKH2011{]}P4  &         & X       & 12:17:02 & 46:54:20 &                    & 20.62   \\
{[}KKH2011{]}P3  &         & X       & 12:17:06 & 46:54:08 &                    & 21.19   \\
dw1220+4748      & M       &         & 12:20:56 & 47:48:59 & 11.3               & 21.61   \\
{[}KKH2011{]} P2 &         & X       & 12:18:06 & 46:28:50 &                    & 22.25   \\
{[}KKH2011{]}S11 & C       & X       & 12:20:30 & 47:29:27 & 9.2                & 23.38   \\
588017110759637262$^{a}$&    & X       & 12:24:11 & 47:07:21 &                    & 18.91   \\
M106edgeN4217    &         &         & 12:16:12 & 47:08:04 &                    & 18.4$^{b}$ \\ \hline
\end{tabular}
}
\vspace*{3ex}

    \begin{minipage}{\textwidth}
NOTE—Column (1): Common names of the galaxy in the literature or online database. $^{a}$ Galaxy ID in SDSS DR7. Column (2): classification of this galaxy in \cite{Carlsten2021ApJ}. C: Confirme; M: possible; N: background. Column (3): classification of this galaxy in \cite{Spencer2014ApJ}. Y: Probable Satellites; X: Possible Satellites. Column (4): R.A.(J2000). Column (5): Dec. (J2000). Column (6): Distance measured by SBF. Column (7): The r-band apparent magnitude of galaxies. $^{b}$ M106edgeN4217 has no r-band information, here we present its b-band value. 
    \end{minipage}
\end{table*}

\begin{landscape}
\begin{table}
\caption{Properties of the satellite galaxies in the M 106 group}
\label{tableA2}
\resizebox{0.86\paperheight}{!}{%
\begin{tabular}{cccccccccccccccccc}
\hline
\hline
(1)              & (2)  & (3)      & (4)      & (5)  & (6)         & (7) & (8)    & (9)  & (10)    & (11)        & (12)          & (13)         & (14)                   & (15)                   & (16)& (17)& (18)\\
Name             & Code & RA       & Dec      & T& V           & D   & Method & ref. & $M_{B}$ & $logD_{25}$ & $logM_{star}$ & $logM_{26}$  & $logSFR_{H\alpha }$    & $logSFR_{FUV}$         & $f_{HI}$        & $logM_{HI}$  & PD     \\
                 &      & (J2000)  & (J2000)  &      & km s$^{-1}$ & Mpc &        &      &         & 0.1arcmin   & $M_{\odot }$  & $M_{\odot }$ & $M_{\odot }$ yr$^{-1}$ & $M_{\odot }$ yr$^{-1}$ & Jy  km s$^{-1}$ & $M_{\odot }$ & kpc    \\ \hline
M106             & 0    & 12:18:58 & 47:18:14 & Sbc& 454         & 7.6 & Maser  & 4    & -20.94  & 2.23        & 10.91& 11.32& 0.43& 0.39& 483.00          & 9.82         & 0.00   \\
NGC 4248/S04     & 1    & 12:17:50 & 47:24:33 & Sm& 483         & 6.8 & TRGB   & 2    & -17.75  & 1.30        & 8.84& 8.91& -1.90& -2.20& 3.20            & 7.64         & 29.09  \\
S05              & 1    & 12:18:11 & 46:55:02 & Sm& 402         & 8.3 & TRGB   & 3    & -13.60  & 0.90        & 7.41&              &                        &                        & 3.10            & 7.63         & 53.24  \\
UGC 7356/S07     & 1    & 12:19:09 & 47:05:24 & Ir& 269         & 7.3 & TRGB   & 5    & -13.32  & 0.95        & 7.74&              & -3.88& -4.76& 3.45            & 7.67         & 28.22  \\
NGC 4288         & 2    & 12:20:38 & 46:17:30 & Sd& 531         & 8.0 & TF     & 6    & -18.90  & 1.23        & 8.56& 9.72& -1.30& -1.17& 47.60           & 8.81         & 139.29 \\
S14              & 2    & 12:20:55 & 46:49:48 &      & 480$^{a}$   & 7.9 & SBF    & 7    & -10.08  & 0.17        & 6.99&              &                        & -4.83& 0.085           & 6.06         & 76.46  \\
NGC 4144         & 3    & 12:09:58 & 46:27:27 & Scd& 265         & 7.2 & TRGB   & 1    & -18.45  & 1.72        & 9.33& 9.84& -0.84& -0.73& 59.00           & 8.90         & 232.43 \\
NGC 4242         & 3    & 12:17:30 & 45:37:09 & Sdm& 516         & 5.3 & TRGB   & 2    & -17.46  & 1.58        & 9.44& 9.85& -1.08& -1.95& 42.00           & 8.76         & 225.26 \\
UGC 7408/DDO 120 & 3    & 12:21:15 & 45:48:43 & Ir& 459         & 7.7 & TRGB   & 1    & -16.50  & 1.30        & 8.76& 8.18& -4.10& -1.90& 8.50            & 8.06         & 204.36 \\
J1211+4739       & 4    & 12:11:35 & 47:39:27 & Ir& 752         & 7.6 & mem    &      & -13.87  & 0.57        &               &              &                        &                        & 0.85            & 7.06         & 171.72 \\
J1215+4730       & 4    & 12:15:52 & 47:30:17 & Ir& 660         & 7.6 & mem    &      & -14.27  & 0.68        &               &              &                        &                        & 0.11            & 6.18         & 74.89  \\
UGC 7391         & 4    & 12:20:16 & 45:54:30 & Scd-Sd& 737         & 7.6 & mem    &      & -15.27  & 1.01        &               &              &                        &                        & 2.95            & 7.60         & 186.81 \\ \hline
\end{tabular}%
}
 \par\medskip\footnotesize
 NOTE—Column (1): Common names of the galaxy in the literature or online database. Column (2): Code of the satellite candidates. The host galaxy M 106 is marked as code 0. Column (3): R.A.(h:m:s, J2000). Column (4): Dec.(d:m:s, J2000). Column (5): The Hubble type using the RC3 morphological classification. Column (6): Mean Heliocentric radial velocity from LEDA in km s$^{-1}$. $^{a}$ The velocity is measured by FAST HI spectral line. Column (7): The distance to the galaxy in Mpc. Column (8):Method used to obtain distance. Column (9):References.1: \cite{Tikhonov2018AstBu}; 2: \cite{Sabbi2018ApJS..235...23S}; 3:\cite{Carlsten2020ApJ}; 4: \cite{Humphreys_2013}; 5: \cite{Tully2013}; 6: \cite{Tully2008ApJ}; 7:\cite{Carlsten2021ApJ} Column (10): The absolute B-band magnitudes (corrected for galactic extinction) of the galaxy from LEDA. Column (11): $D_{25}$, the diameter of the galaxy in the B-band at the 25th mag arcsec$^{-2}$. Column (12): The logarithm of the stellar mass in solar units. Column (13): The logarithm of the dynamic mass in solar units. Column (14): The logarithm of SFR based on the $H\alpha $ flux. Column (15): The logarithm of SFR based on the FUV flux.  
 Column (16): HI flux in Jy km s$^{-1}$ observed by FAST. Column (17): The logarithm of the HI mass in solar units derived from HI flux, assuming that galaxies are at the same distances as distance to M 106. Column (18): Projected distance of the galaxy away from M 106, expressed in kpc.

\end{table}
\afterpage{\clearpage}
\end{landscape}


\bsp	
\label{lastpage}
\end{document}